# The Anomalous Formation of Irradiation Induced Nitrogen-Vacancy Centers in 5-Nanometer-Sized Detonation Nanodiamonds

Frederick T.-K. So[1,2,3], Alexander I. Shames[4], Daiki Terada[2,3], Takuya Genjo[2,3], Hiroki Morishita[1], Izuru Ohki[1], Takeshi Ohshima[5], Shinobu Onoda[5], Hideaki Takashima[6], Shigeki Takeuchi[6], Norikazu Mizuochi[1], Ryuji Igarashi[3,7], Masahiro Shirakawa[2,3]*, Takuya F. Segawa[2,8]*

[1]Institute for Chemical Research, Kyoto University, Gokasho, Uji, Kyoto, 611-0011, Japan.

[2]Department of Molecular Engineering, Graduate School of Engineering, Kyoto University, Nishikyo-Ku, Kyoto 615-8510, Japan

[3]Institute for Quantum Life Science, National Institutes for Quantum Science and Technology, Anagawa 4-9-1, Inage-ku, Chiba 263-8555, Japan

[4]Department of Physics, Ben-Gurion University of the Negev, Israel

[5]Takasaki Advanced Radiation Research Institute, National Institutes for Quantum Science and Technology, 1233 Watanuki, Takasaki, Gunma 370-1292, Japan

[6]Department of Electronic Science and Engineering, Kyoto University, Kyotodaigakukatsura, Nishikyo-ku, Kyoto 615-8510, Japan

[7]National Institute for Radiological Sciences, National Institutes for Quantum Science and Technology, Anagawa 4-9-1, Inage-ku, Chiba 263-8555, Japan

[8]Laboratory for Solid State Physics, ETH Zurich, Otto-Stern-Weg 1, 8093 Zürich, Switzerland

* E-mail: shirakawa@moleng.kyoto-u.ac.jp; takuya.segawa@alumni.ethz.ch

# 5 nm Detonation Nanodiamonds
## as Quantum Sensors

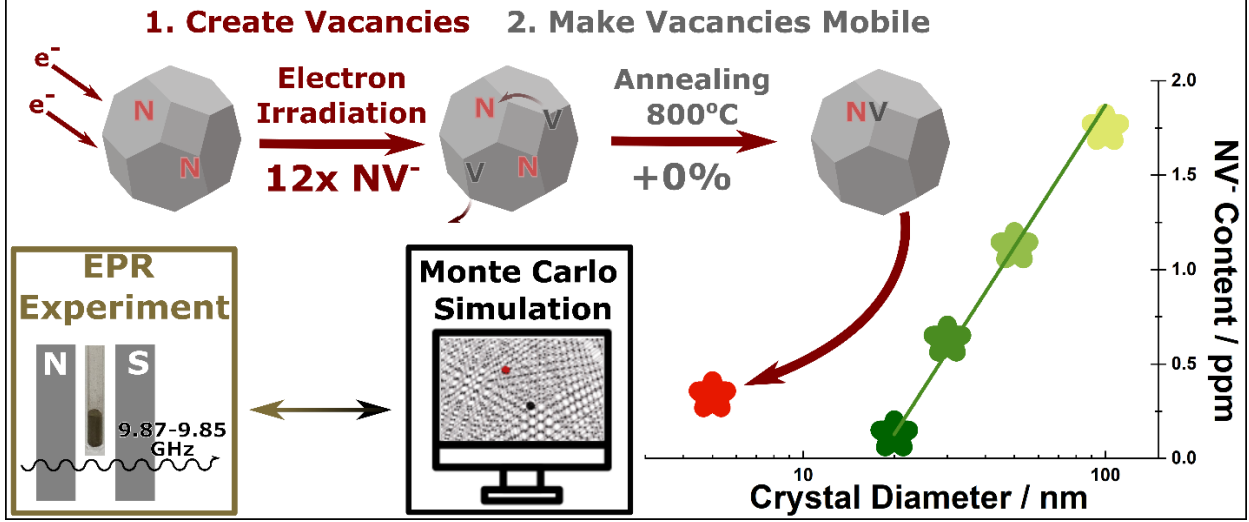

**1. Create Vacancies**   2. Make Vacancies Mobile

e⁻

e⁻

N

N

**Electron Irradiation**

**12x NV⁻**

N   V

V   N

Annealing
800°C

+0%

N V

EPR Experiment

N   S

9.87-9.85 GHz

Monte Carlo Simulation

NV⁻ Content / ppm

2.0

1.5

1.0

0.5

0.0

Crystal Diameter / nm

10       100

**ABSTRACT**

Nanodiamonds containing negatively charged nitrogen-vacancy (NV$^-$) centers are versatile room-temperature quantum sensors in a growing field of research. Yet, knowledge regarding the NV$^-$ formation mechanism in very small particles is still limited. This study focuses on the formation of the smallest NV$^-$-containing diamonds, 5 nm detonation nanodiamonds (DNDs). As a reliable method to quantify NV$^-$ centers in nanodiamonds, half-field signals in electron paramagnetic resonance (EPR) spectroscopy are recorded. By comparing the NV$^-$ concentration with a series of nanodiamonds from high-pressure high-temperature (HPHT) synthesis (10 - 100 nm), it is shown that the formation process in 5 nm DNDs is unique in several aspects. NV$^-$ centers in DNDs are already formed at the stage of electron irradiation, *without* the need for high-temperature annealing. The effect is explained in terms of "self-annealing", where size and type dependent effects enable vacancy migration close to room temperature. Although our experiments show that NV$^-$ concentration generally increases with particle size, remarkably, the NV$^-$ concentration in 5 nm DNDs surpasses that of 20 nm-sized nanodiamonds. Using Monte Carlo simulations, we show that the ten times higher substitutional nitrogen concentration in DNDs compensates the vacancy loss induced by the large relative particle surface. Upon electron irradiation at a fluence of $1.5 \times 10^{19}$ e$^-$/cm$^2$, DNDs show a 12.5-fold increment in the NV$^-$ concentration with no sign of saturation. These findings can be of interest for the creation of defects in other very small semiconductor nanoparticles beyond NV-nanodiamonds as quantum sensors.

KEYWORDS: nitrogen-vacancy center (NV$^-$ center), detonation nanodiamonds (DNDs), electron irradiation, annealing, defect formation in nanocrystals, quantum sensors, electron paramagnetic resonance spectroscopy (EPR spectroscopy)

## INTRODUCTION

Nanodiamonds (NDs) have been an active research topic in recent years due to their versatile applications in various fields such as biophysics,[1] biomedical chemistry,[2,3] quantum sensing,[4] and many more. Among all, quantum sensing applications based on the negatively charged nitrogen-vacancy (NV⁻) centers are the most promising ones due to their specificity, which has its origin in a unique combination of magnetic and optical properties. The NV⁻ center is one out of many fluorescent "color centers" in diamond,[5] containing a substitutional nitrogen defect (a nitrogen atom which replaces a carbon atom in the crystal lattice) next to a neighboring vacancy (a missing carbon atom). This defect absorbs green and emits red to near infrared fluorescence with a zero-phonon line at 637 nm.[6] Two unpaired electrons make the NV⁻ center a paramagnetic defect with a total electron spin $S = 1$ and a characteristic zero-field splitting of $D \approx 2.87$ GHz (originating from the dipole-dipole coupling between the two unpaired electron spins). The two key features of this unique quantum system are the optical polarization of the electron spin transition via the excitation laser combined with the optical detection of the electron spin state via the observation of the fluorescence signal. This enables the detection of a single NV⁻ center at zero magnetic field and room temperature.[7,8] Thanks to the protection of the surrounding diamond crystal lattice, the NV⁻ center shows very long electron spin coherence times.[9] At the same time, the spectral properties of the NV⁻ center are selectively altered by change of external factors, such as magnetic field or temperature.[1] By incorporating NV⁻ centers into diamond nanocrystals (ca. 5-100 nm),[10] quantum sensors for the three-dimensional orientation,[11,12] for temperature[13] or pH[14,15] were presented in the last decade. These are all quantities, which are of high biological relevance, especially for measurements inside living cells, and are currently not yet sufficiently covered by existing fluorescent sensors.[16]

Most NDs are produced via the conventional high-pressure high-temperature (HPHT) synthesis at ambient atmosphere, thus naturally contain a certain amount of nitrogen, occurring predominantly as single substitutional nitrogen defects $N_S$.[17] These are the precursors for NV⁻ centers. In addition, to create vacancies, diamonds are irradiated with high-energy electrons or ions.[18,19] The final step to create NV⁻ centers is high-temperature annealing (usually 800 °C or higher). This initiates the "breaking and making" of carbon-carbon bonds, which effectively leads to a migration of the vacancies. Once a vacancy is placed next to a substitutional nitrogen $N_S$, the energetically more stable NV⁻ center is formed. The annealing procedure is carried out under high vacuum or inert gas to avoid surface oxidation and graphitization.[20] Conventionally, all the above steps for the NV⁻ center creation are carried out on micron-sized diamonds (microdiamonds). In a final step to create nanodiamonds, the microdiamonds are milled down to nanodiamonds, where different size distributions are selected by centrifugation.[21] This is called the "top-down" approach for the synthesis of fluorescent nanodiamonds (FNDs).[22] In contrast, the "bottom-up" approach, as studied in the current work, creates first nano-sized diamond particles, followed by the NV⁻ center creation (irradiation and annealing).

An alternative way to artificially synthesize nanodiamonds is the detonation synthesis.[23] By an explosion of reagents such as trinitrotoluene (TNT) and hexogen in a reaction chamber, the shock wave transforms the carbon atoms from these explosive molecules into nanodiamonds. While pioneering experiments in the USSR date back to the 1960's, the deaggregation of these very strongly aggregated nanodiamonds succeeded only in the beginning of the 21st century.[24] The isolated particles have a rather uniform size of 4-5 nm, which makes detonation nanodiamonds (DNDs) to the smallest type of NDs that are capable of hosting color defects,[25–27] which can be produced in bulk quantities. In contrast to milled HPHT NDs, which often have irregular flake-like shapes,[28,29] DNDs have uniform shapes, which are close to spheres (truncated octahedra).[30] The content of nitrogen in DNDs, measured by elemental analysis reaches 2-3 at.%,[31] which is by far more than in any other type of diamond. For instance, most of as-synthesized commercial HPHT micron-sized diamonds contain about $100 - 200$ ppm of nitrogen thus belonging to the class of type Ib diamond.

In contrast to all other nano- and bulk diamonds, a substantial concentration of $NV^-$ centers already exists in "pristine" DNDs, which were not treated by any post-synthesis irradiation and/or high-temperature annealing.[25,32,33] Recently, we have shown that this basic $NV^-$ concentration can be further enhanced through electron irradiation.[25] To our surprise, we observed that $NV^-$ centers were already created after electron irradiation without the need for the subsequent high-temperature annealing step. Again, this is a behavior, which is unique to DNDs (and, supposedly, to other types of NDs fabricated by dynamic synthesis)[33] and not seen in nano- or in bulk diamonds obtained by static (or quasi-static, like CVD) synthesis. An exception is the recently reported femtosecond laser irradiation technique for the creation of $NV^-$ centers in bulk diamond, where post-irradiation annealing is not necessary to mobilize vacancies in $NV^-$ formation.[34] However, the femtosecond laser method arguably follows a different mechanism of vacancy creation[35] as well as induces high amount of heat that may cause annealing during irradiation. Hence, this method will not be further discussed in this paper.

The goal of this study is to shed light behind these anomalous effects in the formation of $NV^-$ centers in DNDs. Is it the very small size of DNDs, which is responsible for these effects? Or is it rather the different class of NDs characterized by a huge concentration of nitrogen and many other impurities and defects? Why are $NV^-$ centers in DNDs created during electron irradiation at temperature between room temperature and maximum 100 ˚C, in contrast to all other nano- and bulk diamonds? Is there a size-dependent heating effect, which only plays for 5 nm DNDs?[36] To answer these questions, we compare the NV[-] concentration in a series of NDs with different sizes. Besides 5 nm DNDs, the sample of our interest, we examined commercial HPHT NDs with sizes of 10 nm, 20 nm, 30 nm, 50 nm, and 100 nm. NV[-] concentration was measured in (1) their pristine form (no irradiation or annealing), (2) after electron irradiation only and (3) after subsequent conventional high-temperature annealing (800 °C at 2 h). For DNDs and 100 nm nanodiamonds, we also compared the NV[-] concentration as a function of the electron fluence (up to $1.5 \times 10^{19}$ e[-]/cm²), since we are interested to know, to which extent NV[-] centers can be enriched in DNDs. These experiments are repeated at two different electron kinetic energies (2 MeV and 1 MeV, respectively) to test, if heating effects during irradiation may affect the NV[-] formation. For DNDs, we also scan the annealing temperature from 400 to 800 °C, to test, if the optimal annealing temperature could be lower for these 5 nm-sized diamond particles. At the same time, we observe the creation and annihilation of other paramagnetic defects in nanodiamonds, notably the negatively charged vacancies V[-] in diamond. Finally, we monitor DND's crystal integrity using transmission electron microscopy (TEM) and electron energy loss spectroscopy (EELS), to confirm that prolonged electron irradiation does not cause significant damage to the lattice.

In this study, we use continuous-wave Electron Paramagnetic Resonance (EPR) spectroscopy to determine the concentration of the NV[-] centers in NDs. To the best of our knowledge, this is currently the only reliable technique, which can assign and quantify NV[-] centers in powders of nanodiamonds with sizes down to 5 nm.[37] We use the so-called half-field (HF) transitions, which appear approximately at half the magnetic field (i.e., at an effective $g$ factor $g_{eff}$ of ca. 4 or larger), compared to the resonant magnetic field for free $S = 1/2$ electron spins having a $g$-value of ca. 2.[38,39] The HF transition is a forbidden transition with a change in the spin quantum number "$\Delta M_S$

= 2" for $S = 1$ electron spins (such as the NV⁻ center), where a transition between superpositions of $M_S = -1$ to $M_S = +1$ (and vice versa) becomes possible. Such a state mixing occurs, when the zero-field splitting parameter $D$ is comparable to the Zeeman splitting energy. The resonance of the HF signal characteristically depends on the zero-field splitting (ca. 2.87 GHz for NV⁻ centers) and enables an unambiguous assignment of the triplet defect, with an effective $g$ factor $g_{eff} = 4.23$ in this work, where $g_{eff}$ is the effective $g$ parameter, $g_{eff} = h\nu/(\beta_e B)$, where $h$ is the Plank's constant, $\nu$ is the microwave resonant frequency ($\nu = 9.87^*$ GHz), $\beta_e$ is the Bohr magneton, $B$ is the magnetic field applied. While the allowed $\Delta M_S = 1$ transitions are not detectable in a powder of small NDs due to the strong orientation dependence, fortunately, the HF transition is only weakly affected by the orientation dependence of the NV⁻ centers in a nanodiamond powder.[40] For a compact introduction of this approach, the HF EPR method to detect triplet centers in nanodiamond powders was recently reviewed in a viewpoint article.[37]

## EXPERIMENTAL RESULTS

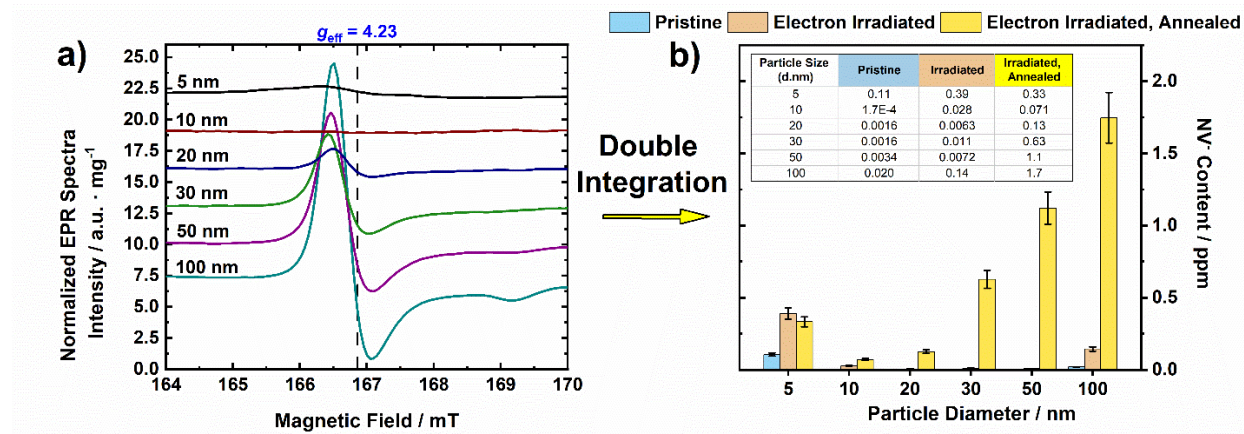

Figure 1. (a) Continuous-wave HF EPR spectra ($\nu$ = 9.87* GHz) of electron-irradiated (2 MeV, 5×10$^{18}$ e⁻/cm²) nanodiamonds with different particle sizes after high-temperature annealing. The double integral of the signal at $g_{eff}$ = 4.23 provides the NV⁻ concentration. (b) A summary graph of NV⁻ content (ppm, in units of atomic ratio) in NDs of different sizes, measured via the HF EPR technique. Blue bars are the pristine nanodiamonds, orange bars are the electron-irradiated nanodiamonds and yellow bars are the electron-irradiated and subsequently annealed nanodiamonds (derived from the $g_{eff}$ = 4.23 signal in the EPR spectrum of Fig. 1(a)). Electron irradiation was conducted with 2 MeV electrons at a fluence of 5×10$^{18}$ e⁻/cm². Annealing was performed at 800 °C in vacuum for 2 h. All samples were boiling acid treated[25] at 130 °C for 3 days to remove Fe$^{3+}$ impurities, which overlap with the HF EPR NV⁻ signal.[38] Inset table shows the corresponding NV⁻ concentrations in ppm. Errors in NV⁻ concentration do not exceed ±15%.

Fig. 1(a) shows a series of HF EPR spectra for NDs of different sizes, which were electron-irradiated (2 MeV, 5×10$^{18}$ e⁻/cm²) and annealed at 800 °C in vacuum for 2 h. For all samples, the characteristic NV⁻ HF transition appears at ~166.9 mT ($g_{eff}$ = 4.23 at a microwave frequency of around 9.87 GHz, X-band). One can see that $g_{eff}$ of the HF line does not depend on the particle size and the signal gets more intense for larger ND particle sizes. In addition, DNDs have a distinctively broader linewidth than larger nanodiamonds.[38] Fig. 1(b) shows the overview of the results for the DNDs and the size series of HPHT nanodiamonds. All nanodiamonds were measured in their pristine ("as-received") form (blue bars), after electron irradiation with 2 MeV electrons at a fluence of $5 \times 10^{18}$ e$^-$/cm$^2$ (orange bars) and after additional high-temperature annealing at 800 °C in vacuum for 2 h (yellow bars). All samples were boiling acid treated before the EPR measurement to remove $Fe^{3+}$ impurities,[25] where their broad signal overlap with the HF EPR NV$^-$ signal.[38] About 20 mg of DND powder was filled into the EPR tube and the NV$^-$ content was measured by taking the double integral of the corresponding HF EPR signal ($g_{eff}$ = 4.23), which was normalized to the weight and compared to a reference sample (see Materials and Methods for more detail).

DNDs clearly stand out of this series: (1) Pristine DNDs show a substantial NV$^-$ concentration, which is not seen for any of the pristine HPHT nanodiamonds. (2) Electron irradiation alone, *without* high-temperature annealing, creates a substantial additional amount of NV$^-$ centers in DNDs. The subsequent high-temperature annealing does not create further NV$^-$ centers in DNDs. Again, this is in strong contrast to any of the HPHT nanodiamonds, where only the final step of high-temperature annealing creates the main amount of NV$^-$ centers. (3) The concentration of the NV$^-$ centers in DNDs is very high, higher than for 10 nm and 20 nm HPHT nanodiamonds. All concentrations are given in units of atomic ratio, i.e., the number of NV$^-$ centers, normalized by the total number of carbon atoms in a nanodiamond. For pristine DNDs, we estimated that 1 out of ca. 1000 DNDs contain an NV$^-$ center, while for $5 \times 10^{18}$ e$^-$/cm$^2$ electron irradiated DNDs, we estimated that 1 out of ca. 250 DNDs should contain an NV$^-$ center.[25] For irradiated and annealed 10 nm HPHT NDs, despite their 8 times larger volume, we estimate that only 1 out of ca. 180 NDs contain an NV⁻ center. Finally, for the size series of HPHT NDs, we see a clear increase of NV⁻ concentration (ppm, in atomic ratio) as a function of particle size (see yellow bars in Fig. 1(b)).

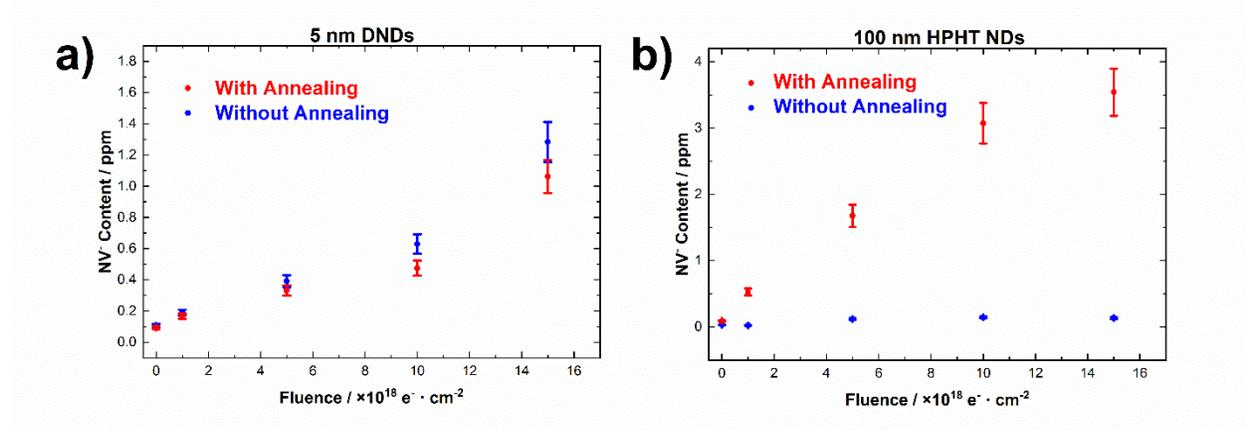

Figure 2. Electron fluence dependence of NV⁻ concentration for (a) DNDs and (b) 100 nm HPHT NDs. Blue dots are before annealing and red dots are after annealing (800 °C in vacuum for 2 h). Errors in NV⁻ concentration do not exceed ±15%. Data points at 0×, 1×, 5×$10^{18}$ e⁻/cm² for DNDs in Fig. 2 (a) were derived from EPR spectra published by the authors in Ref. 25.

The electron fluence dependence of the NV⁻ concentration in DNDs and 100 nm HPHT NDs, respectively, were measured in a range between 0 to 1.5×$10^{19}$ e⁻/cm². The NV⁻ concentration was determined before (blue) and after (red) high-temperature annealing (800 °C in vacuum for 2 h) and is plotted in Fig. 2. While DNDs show an approximately linear increase of NV⁻ concentration with increasing fluence, for 100 nm HPHT NDs a saturation of the NV⁻ concentration after 1×$10^{19}$ e⁻/cm² becomes visible, as it was previously described for type Ib microdiamonds.[41] As already seen in Fig. 1, the NV⁻ concentration of 100 nm HPHT NDs without high-temperature annealing stays close to zero, while for DNDs, the NV⁻ concentration consistently increases through electron irradiation only, *without* the need for annealing. The highest NV⁻ concentrations in this study were reached at the fluence of $1.5 \times 10^{19}$ e⁻/cm²: DNDs showed an NV⁻ concentration of 1.3 ppm (before annealing), while 100 nm HPHT NDs had an NV⁻ concentration of 3.5 ppm (after annealing), which is roughly 2.7 times higher. This corresponds to 1 out of ca. 80 DNDs containing an NV⁻ center versus a single 100 nm HPHT ND containing about 280 NV⁻ centers on average.

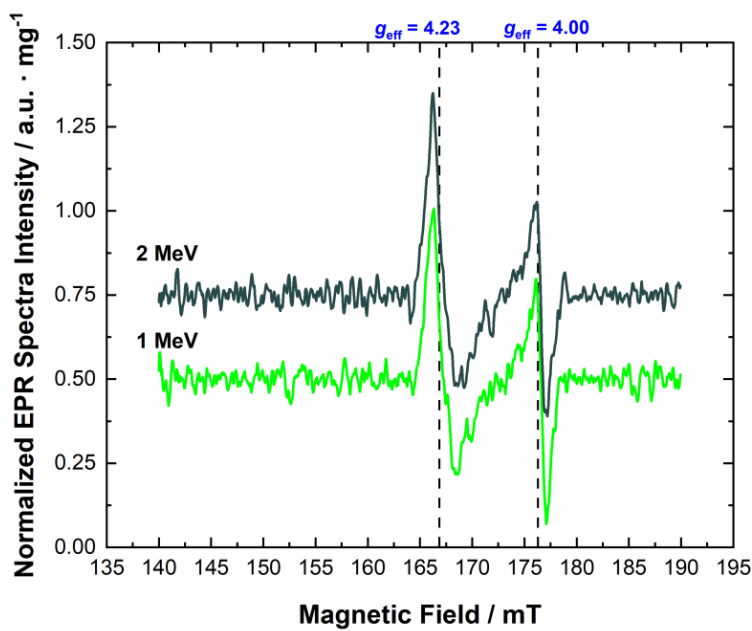

Figure 3. HF EPR spectra (ν = 9.87* GHz) of electron irradiated DNDs with 2 MeV electrons (dark green) and 1 MeV electrons (light green) at a fluence of $5 \times 10^{18}$ e⁻/cm². The spectra shown were background corrected (see Materials and Methods for details).

"Self-annealing" is one of our hypotheses explaining why additional NV⁻ centers in 5 nm DNDs are already formed during electron irradiation. Related either to the very small size or the special diamond type of DNDs, this could be responsible for a room-temperature annealing process, which would take place for vacancies in DNDs, but not for vacancies in all HPHT NDs. One such effect would be a local and size-selective heating of the very small DND particles, due to their lower heat capacity and lower thermal conductivity in comparison to larger diamond particles. As an experimental test of this hypothesis, electron irradiation ($5\times10^{18}$ e$^-$/cm$^2$) was carried out for 2 MeV and 1 MeV electrons, respectively. While the creation of a vacancy in diamond "costs" only a few electron volts, a part of the electron kinetic energy will be converted into heat, when hitting a DND particle. If the experimental results would show a smaller NV$^-$ concentration in DNDs for 1 MeV electrons than 2 MeV, this would support the "self-annealing" hypothesis. As can be seen for Fig. 3, the HF EPR spectra for DNDs irradiated with 2 MeV and 1 MeV electrons look identical (NV$^-$ HF signal at $g_{\text{eff}}$ = 4.23). In the energy range between 1 MeV to 2 MeV, we do not see any difference, which could support the self-annealing hypothesis. The second signal at $g_{\text{eff}}$ = 4.00 is attributed to multi-vacancies or coupled dangling bonds, which are always present in HF EPR spectra of DNDs.[33,39]

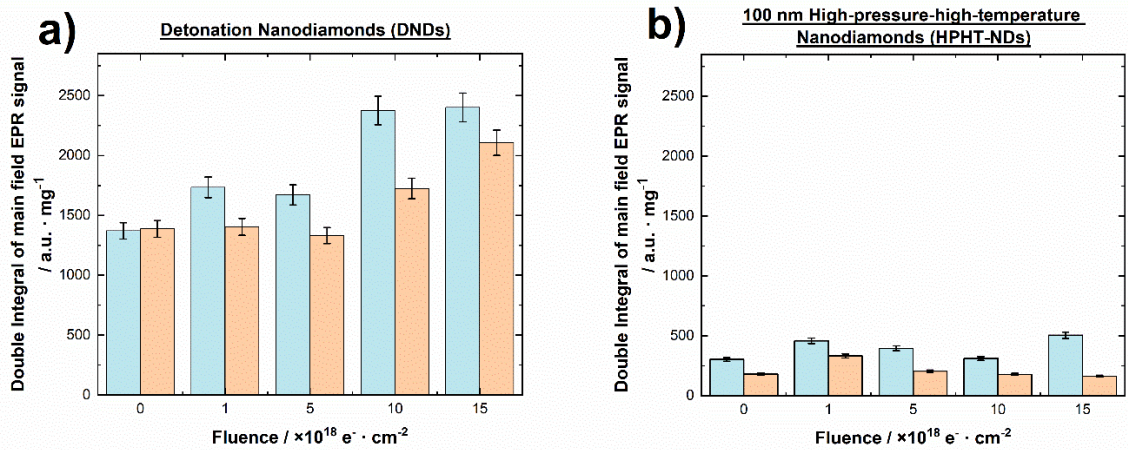

Figure 4. Double integral of the EPR line at $g \approx 2$ ("main field signal") for different electron irradiation fluences before (blue) and after (orange) annealing in a) 5 nm DND and b) 100 nm HPHT NDs. Paramagnetic species contributing to these signals are P1 centers ($S$ = ½), isolated

dangling bonds ($S = \frac{1}{2}$) and negatively charged vacancies V⁻ ($S = 3/2$). Errors in the double integral value do not exceed ±5%.

Measurements of the EPR signal at $g \approx 2$ are not as selective as HF EPR signals and include all $S = \frac{1}{2}$ electron spin signals, such as neutral substitutional nitrogen defects $N_S^0$ (called P1 centers in the EPR literature),[39] isolated dangling bonds as well as irradiation induced negatively charged vacancies V⁻ having $S = 3/2$.[42] It is well known that the total concentration of $S = \frac{1}{2}$ electron spins in DNDs is very high compared to larger diamond particles.[38] In case of pristine, non-irradiated, nanodiamonds, we can assume the absence of negatively charged vacancies V⁻ and determine the total concentration of $S = \frac{1}{2}$ spins. For pristine DNDs, a total $S = \frac{1}{2}$ spin concentration of 1280 ± 220 ppm (average over three DND samples) was obtained, while for pristine 100 nm HPHT NDs a total S = ½ spin concentration of 106 ppm was measured (see SI for details), where the latter value is in good agreement with recent experiments in microdiamonds.[41] By decomposing the $g \approx 2$ signal of the non-irradiated DNDs into two Lorentzian components,[43] the P1 signal can be assigned to the component with the narrow linewidth.[44,45] Using this approach, the P1 content in 5 nm DNDs was estimated to be 964 ± 110 ppm (see SI for details). In a general trend, electron irradiation at increasing fluence also increases the $g \approx 2$ signal intensity for both types of nanodiamonds (see Fig. 4). High-temperature annealing consistently decreases this signal, with the notable exception of non-irradiated DNDs, where the $g \approx 2$ EPR peak is unaffected by high-temperature annealing (Fig. 4). For DNDs, the reduction in percentage of the EPR double integral varies in a range of 0% to 30%, for 100 nm HPHT NDs in a range of 30 to 70%. The drop of the $g \approx 2$ EPR double integral is mainly assigned to the loss of negatively charged vacancies V⁻ through annealing. While in 100 nm HPHT NDs, a part of these vacancies form NV⁻ centers, in 5 nm DNDs, they do *not* contribute to an increase of NV⁻ centers (see Fig. 2).

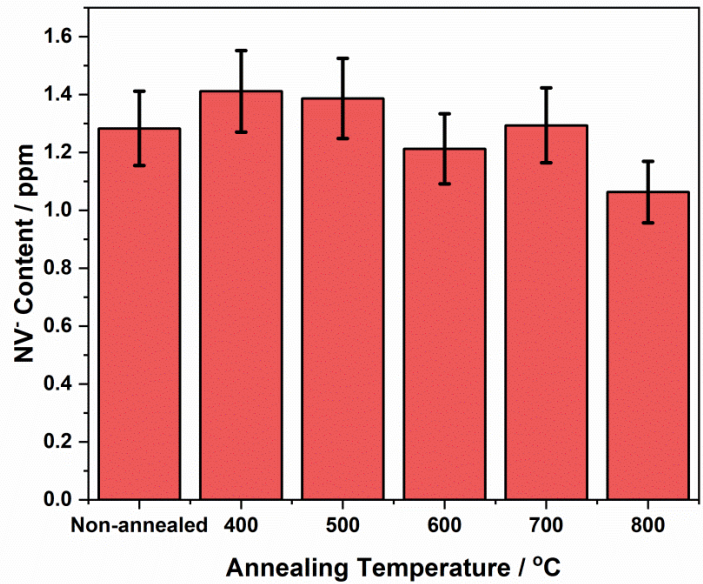

Figure 5. Summary graph of NV⁻ content of $1.5 \times 10^{19}$ e⁻/cm² electron-irradiated DNDs with different annealing temperature, including non-annealed and annealing temperature ranging from 400 °C to 800 °C for 2 h.[46] Errors in NV⁻ concentration do not exceed ±15%.

Conventional annealing temperatures to create NV⁻ centers are 800 °C or higher. To understand whether for DNDs an optimal annealing temperature may exist within a lower temperature range, we performed an annealing with temperatures ranging from 400 °C to 800 °C for 2 h.[46] No significant change in the NV⁻ concentration of electron-irradiated DNDs ($1.5 \times 10^{19}$ e⁻/cm²) was found for annealing at temperatures ranging from 400 °C to 800 °C for 2 h (see Fig. 5). Within error, all concentrations remained at the same level as the irradiated but non-annealed sample.

**DISCUSSION**

**Factors affecting NV⁻ center formation**

**(i) ND particles' size dependence**

With the notable exception of DNDs, Fig. 1 shows a steady increase of NV⁻ concentration in irradiated and annealed NDs on increasing average particle sizes. All NDs were treated identically, by irradiation with 2 MeV electrons at a fluence of $5 \times 10^{18}$ e⁻/cm² and following high-temperature annealing at 800 °C in vacuum for 2 h. Such a size dependence of NDs for the creation of NV centers was discussed in Ref. 47. Smith et al. compared 5 nm DNDs with 55 nm HPHT NDs by photoluminescence (PL) measurements. By selecting the signal with a long PL lifetime of 12 ns, the intensity of the 55 nm HPHT ND signal was ca. 50 times stronger than for the 5 nm DNDs (studying samples with the same weight equally distributed on a quartz cover slip). In that study, the NV⁻ formation protocol was identical for both types of NDs ($5 \times 10^{15}$ protons/cm² with 2.5 MeV, followed by annealing at 750 °C in vacuo at 1 mPa for 2 h). The authors attributed this difference to loss of vacancies via the ND surface, which accentuates for very small particles due to their large surface-to-volume ratio. Using Monte Carlo simulations, where the vacancy annealing process was simulated by a random walk, two regimes were found (Fig. 6): (1) for very small NDs, the probability to form an N-V pair during annealing increases with the square of its radius $r^2$, since the average number of steps in the random walk of a vacancy required to reach the surface is proportional to $r^2$; (2) for larger NDs, there is a saturation in the NV⁻ formation probability, since a vacancy cannot form more than one NV⁻ center.[47] Fig. 6 shows the results of the Monte Carlo simulation in Ref. 47, where blue dots are simulated NV⁻ formation probabilities, the green line is a fit proportional to $r^2$ and the light blue line is fitting an exponential buildup. To compare the simulation data from Ref. 47 with our experimental values, both data were plotted as a function of the particle diameter (originally radius) using a logarithmic scale in Fig. 6.

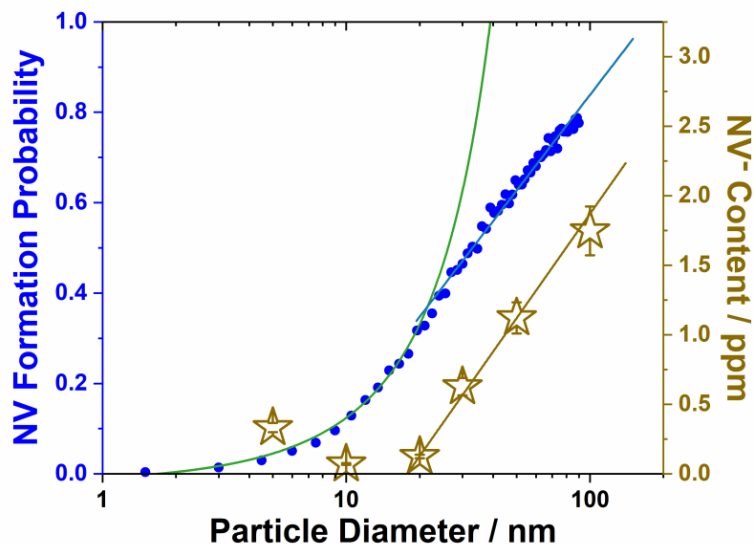

Figure 6. Monte Carlo simulation (left axis, blue circles) for the probability of vacancies forming NV centers in small nanodiamonds with a substitutional nitrogen concentration [$N_S$] of 300 ppm (adapted from Ref. 47), and experimental NV⁻ content (right axis, brown stars) of electron irradiated ($5 \times 10^{18}$ e⁻/cm²) and annealed (800 °C for 2 h) NDs (this work, same data as yellow bars in Fig. 1), plotted as a function of their particle size in logarithmic scale. For small nanodiamonds, the probability of forming NV⁻ centers decreases with the square of the particle radius (fitting curve in green). For relatively large crystals or diamonds the probability saturates (light blue line for simulated data and brown line for experimental data). The saturation behavior starting from a particle diameter of ca. 20 nm is clearly visible for simulated and experimental data, as a linear trend in this semi-logarithmic plot (corresponding to an exponential buildup). The experimental data of this work is vertically shifted with respect to the simulated data of Ref. 47

due to a higher concentration of substitutional nitrogen $N_S$ of 300 ppm in the simulation compared to ca. 70 - 100 ppm for the series of HPHT nanodiamonds in this work.

Indeed, the four largest NDs in the present experimental series (20 nm, 30 nm, 50 nm and 100 nm HPHT NDs, substitutional nitrogen $[N_s]$ concentration estimated to be 70 – 100 ppm) lie, within error, on a linear line, corresponding to an exponential buildup ("saturation") of the NV⁻ formation. The saturation regime in the simulation of Ref. 47 also starts at about a particle diameter of 20 nm. The 10 nm HPHT ND from the experimental series is clearly above this regime, what suggests that NV⁻ concentration is still increasing with radius $r^2$. As a result, the experimental data confirmed the existence of the two formation probability regimes, which, to the best of our knowledge, is the first experimental proof of the Monte Carlo simulation in Ref. 47. As it will be discussed in detail below, the DND has a larger NV⁻ concentration than 10 nm or 20 nm HPHT NDs, despite its small size and does not fit into this trend.

**(ii) Concentration of substitutional nitrogen $N_S$**

What are further parameters, besides particle size, which influence the NV⁻ formation probability in nanodiamonds? We identify two dependent factors as the concentrations of the two "reactants": the vacancies [V] and the substitutional nitrogen atoms $[N_S]$, the two precursors for NV⁻ centers. Since the irradiation conditions are identical for the data in Fig. 1, we suppose that the high NV⁻ formation probability in 5 nm DNDs is caused by the higher concentration of substitutional nitrogen atoms $[N_S]$ in DNDs than in HPHT nanodiamonds. For our HPHT NDs, we estimate a substitutional nitrogen concentration $[N_S]$ of 70 – 100 ppm, based on double integration of the main field EPR signal ($g \approx 2$). Even though DNDs have nitrogen concentration of 2-3 at.%,[31] their substitutional nitrogen concentration $[N_S]$ was estimated to be around 850-1100 ppm (details of the Ns content estimation are presented in SI). The Ns concentration in DNDs is on one hand an order of magnitude higher than in HPHT NDs. On the other hand, it appears remarkably low in relation to the total nitrogen content, since only about 5% appears to be in the form of P1 centers. It is important mentioning here that EPR spectroscopy can only detect one charge state namely the neutral $N_S^0$ (paramagnetic P1 centers with $S = \frac{1}{2}$), but neither the positively-charged substitutional nitrogen atoms $N_S^+$ nor the nitrogen A- and B-defects (different form of nitrogen clusters, all diamagnetic defects with $S = 0$).[18] Other forms of nitrogen in DNDs could be larger "nitrogen-vacancy" clusters such as so-called H3 or H4 defects,[18] however, previous PL and optical absorption measurements did not succeed in detecting theses defects in DNDs.[48] To account for the additional variable of the substitutional nitrogen concentration [Ns] , we performed Monte Carlo annealing simulations by varying both parameters, the particle diameter and [Ns]. Fig. 7 shows results for particles with diameters between 3 nm and 10 nm and substitutional nitrogen concentration [Ns] between 70 and 1000 ppm (i.e., 0.1 at.%). Indeed, the simulations show that, assuming Ns concentration ranges of 900 – 1000 ppm for 5 nm DNDs and 70 – 100 ppm for 10 nm HPHT NDs, the NV⁻ concentration for the 5 nm DNDs will surpass 10 nm HPHT NDs by a factor of about three. In contrast, if for both particle sizes the same concentration [Ns] = 100 ppm is assumed, the 10 nm particles form about four times more NV⁻ centers in the simulation, which is in agreement with the "square of the radius $r^2$ law" described in Ref. 47. The Monte Carlo simulation takes place in a spherical diamond particle, where, for every run, nitrogen atoms of the given concentration are randomly placed in the crystal lattice. Then, a single vacancy is created at a random lattice point. This moves now randomly in one of the four possible directions, from lattice site to lattice. The simulation run ends either with the formation of an NV center, when the vacancy arrives next to nitrogen or with the loss of a vacancy to the surface, when arriving the edge of the crystal. The protocol is repeated, until the probability converges ($10^4$ runs). The simulation is based on a simplified NV$^-$ formation model, where the presence of other defects, which could capture vacancies, such as self-interstitial,[49] or influence the charge equilibrium, like boron acceptors,[50] is ignored. Neither, we do discriminate the charge states of negative and neutral vacancies, where the former one is dominating in type Ib diamond, but the latter one is the charge state during migration.[18] And, finally, we do not include any activation energies for the migration or assume a temperature of the system. Still, in relative terms, this should account for the NV$^-$ formation dependence on ND size and $N_S$ concentration. By assuming spherical crystal shapes, we might be overestimating the volume of the 10 nm HPHT nanodiamonds. While DNDs are of spherical (or truncated octahedron)[30] shape, milled HPHT nanodiamonds are often rather "flake-like" with a larger aspect ratio.[28,29] This could be one reason, why in our simulation, the 5 nm DNDs showed an approximately three times higher concentration in NV$^-$ compared to 10 nm NDs, where in our experimental results, DNDs had about a five to six times higher concentration in NV$^-$ centers than 10 nm HPHT NDs (Fig. 1).

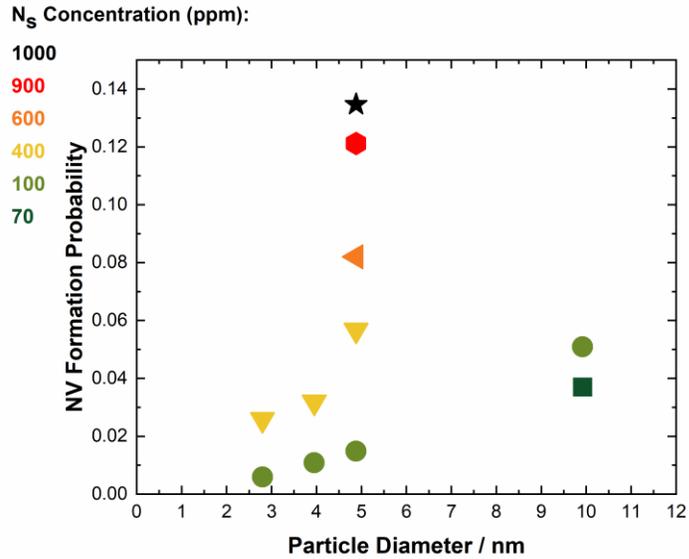

Figure 7. Results of Monte Carlo annealing simulation for the probability of NV center formation as a function of particle diameter and concentration of substitutional nitrogen [$N_S$], ranging from 70 ppm (dark green square), 100 ppm (light green circles), 400 ppm (yellow down-pointing triangle), 600 ppm (orange left-pointing triangle), 900 ppm (red hexagon) and 1000 ppm (black star). NDs are spherical particles with the given diameter in nm. The NV formation probability is the chance that a single vacancy, on a random walk in the crystal lattice, will end up next to a substitutional nitrogen atom $N_S$.

### (iii) Room-temperature annealing in DNDs

How can be explained that, in contrast to all other types of NDs and bulk diamonds, NV$^-$ centers in 5 nm DNDs are formed during room-temperature electron irradiation (maximum temperature 100 °C), with no subsequent high-temperature annealing? Initially, one of our assumptions was that DNDs have so many nitrogen impurities that NV$^-$ centers could be formed "by chance", where the vacancy is directly created next to a substitutional nitrogen atom $N_s$, without the need to further move inside the ND lattice. Then, high-temperature annealing seems to be ineffective, since 5 nm

DNDs are so small that all vacancies would be lost to the surface. However, knowing that the concentration of substitutional nitrogen defects $N_s$ does not exceed 1000 ppm (i.e., 0.1 at.%), the above working hypothesis may be abandoned. As a second hypothesis, we assume that the annealing process for DNDs has already taken place in the course of the electron irradiation (so-called "self-annealing"). The activation energy of vacancy migration ("diffusion barrier") of the neutral vacancy $V^0$ is $E_a \approx 2.1 - 2.3$ eV (the vacancy is considered not to be mobile, before ionization), corresponding to a temperature of ca. 600 °C.[18,51] The necessary diffusion length in 5 nm DNDs is much smaller than in larger particles. Combined with the higher concentration of substitutional nitrogen [$N_S$], this would allow for a lower annealing temperature to create NV⁻ centers in DNDs. To test, if annealing temperatures lower than 800 °C can be effective for DNDs, an annealing temperature series was performed and the results were summarized in Fig. 5. The NV⁻ concentration was independent of the annealing temperature in a range of 400 - 800 °C and identical to the non-annealed sample (all DNDs previously irradiated with a fluence of $1.5 \times 10^{19}$ e⁻/cm²). This result is a further evidence indicating that annealing has already been completed during the electron irradiation process and the NV⁻ concentration cannot be further increased by high-temperature annealing.

What could be size dependent effects in diamond, which play a role in this "self-annealing" process during electron irradiation? Among a variety of the outstanding physical properties, diamond is a material with a very high thermal conductivity of above 2000 W/(m·K). However, this value decreases drastically for particle sizes below 1 μm. Could this cause overheating of small nanoparticles? In Ref. 52, thermal conductivity of diamond nanowires was calculated as a function of their diameter. They derived a remarkably narrow phonon mean free path distribution concentrated at about 1 μm, much larger than the size of all of our NDs in this study. In another recent report, the thermal conductivity of spherical NDs with a diameter of 2.5 nm was calculated to be 10-28 W/(m·K),[53] about two orders of magnitude lower compared to values in bulk diamond. Still, it would be surprising to see an abrupt change of this property by just halving the diameter of a particle from 10 nm to 5 nm. In other words, if only this effect would play a role, we would expect to see, at least some "self-annealing" in 10 nm or maybe even in 20 nm HPHT nanodiamonds, which can be excluded in our experimental data (Fig. 1). Another size dependent parameter related to temperature is the heat capacity $c$ which is given as $c = V\sigma_c$, where $V$ is the volume and $\sigma_c$ is the specific heat capacity per unit volume with $\sigma_c = 1.78 \times 10^6 J/m^3 /K$ for bulk diamonds.[36] The steady state temperature rise is given as:

$$\Delta T = \frac{E_{avg}}{V \sigma_c}$$

where $E_{avg}$ is the average energy dissipated during the heating period. It is difficult to estimate relevant numbers for NDs under electron irradiation, but the temperature rise $\Delta T$ is inversely proportional to the volume V of the particle, which corresponds to about one order of magnitude from 5 nm to 10 nm NDs. To test, if local heating of the very small ND particles was caused by electron irradiation, we repeated the irradiation experiments applying half the electron kinetic energy, i.e., 1 MeV instead of the usual 2 MeV. If the electron irradiation energy was a heating source, which induces annealing, one would expect to see less NV⁻ centers formed in the case of irradiation with 1 MeV electrons. However, as shown in Fig. 3, the concentration of NV⁻ centers formed in DNDs were identical. This is attributed to the very close mass stopping powers of 1.507 MeV·cm²/g for 1 MeV and 1.567 MeV·cm²/g for 2 MeV.[54] 1 MeV is the lowest energy applicable in our radiation facility, corresponding to a penetration depth of 1.2 mm (a minimum of 175 keV is necessary to displace a carbon atom from lattice site). Moreover, since the electron energy is connected to the depth-dose distribution, smaller irradiation energies are less convenient to irradiate larger ND powder quantities. Still, this does not rule out the hypothesis of "self-annealing", since the difference in the deposited energy is small.

If annealing occurs during the electron irradiation at below 100 °C, this could be beneficial for the NV⁻ formation, since a single-step high-temperature irradiation/annealing leads to a higher yield of NV⁻ centers than a two-step protocol with irradiation at room temperature followed by high-temperature annealing for bulk diamonds (ca. factor of two).[55] In the authors understanding, the instantaneous mobilization of vacancies upon creation helps to prevent vacancy aggregation, therefore leading to a more efficient NV⁻ creation. This is another effect, which may contribute to a higher NV⁻ concentration in DNDs through "self-annealing", compared to HPHT nanodiamonds through two-step irradiation/annealing.

A theoretical study predicted a five-fold reduction of the activation energy of vacancy migration $E_a$ of the neutral vacancy near the (001) surface ($E_a = 0.42$ eV).[56] Since the activation energy $E_a$ enters into the exponent of the Arrhenius equation, such a drastic reduction would enable vacancy mobilization and annealing at room temperature: only duration of 1 s would be enough to reach the diffusion length of 5 nm. However, since so far, no experimental evidence for this prediction was reported, it is difficult to judge, if such an effect is also playing a role for the anomalous annealing of 5 nm DNDs.

**(iv) Electron-beam fluence dependence**

As it is clearly seen in Fig. 2a, up to the fluence of $1.5 \times 0^{19}$ e⁻/cm², no saturation has been observed for the NV⁻ center formation in DNDs. This is in contrast to 100 nm HPHT NDs displayed in Fig.

2b, where the NV$^-$ center formation is showing saturation, as it was recently described in similar HPHT microdiamonds.[41] This is consistent with the higher concentration of substitutional nitrogen defects $N_S$ in DNDs, as well as the "ineffective consumption" of vacancies in DNDs, where the vast majority of them gets lost to the surface. This experimental data confirms a deficiency of vacancies with respect to nitrogen impurities in the diamond core of DNDs.[25] In other words, an upper limit for the NV$^-$ concentration in DNDs is not yet in sight, while the diamond crystal structure is safely retained at a fluence of $1.5 \times 10^{19}$ e$^-$/cm$^2$ (see Fig. S7). We show that NV$^-$ centers can be enriched from ca. 1 NV$^-$ center in 1000 DND particles (pristine DNDs) to ca. 1 NV$^-$ center in 80 DND particles (fluence of $1.5 \times 10^{19}$ e$^-$/cm$^2$, before annealing).

**(v) Reduction of negatively charged vacancies V$^-$ during high-temperature annealing in DNDs**

Fig. 4 demonstrates that high-temperature annealing reduces the intensity of the EPR line at $g \approx 2$ for DNDs as well as for 100 nm HPHT NDs. Given that no graphitization was observed in the diamond structure of DNDs (Fig. S7), the reduction of $g \approx 2$ can therefore be largely assigned to the negatively charged vacancies V$^-$, since other defects (P1, dangling bonds) should remain almost unaffected at this temperature. This result is expected, however, for DNDs, it uncovers one more surprise: although vacancies are made mobile, high-temperature annealing *does not form* any new NV$^-$ centers. A similar effect was observed after a second round of annealing at 1400 °C within microdiamonds from HPHT synthesis: a substantial reduction of the negatively charged vacancies V$^-$ by 50 % was accompanied by a slight reduction of P1 centers and also NV$^-$ centers.[41] Indeed, a careful look at Fig. 2a reveals a slight reduction of the NV$^-$ centers induced by annealing, which remains mostly within error, but appears consistently for all fluences. These observations can currently not yet be explained within the frame of our hypothesis. Remarkably, for the non-irradiated DNDs, annealing has no effect on the $g \approx 2$ signal. This suggests that all vacancies created during the detonation synthesis (otherwise, there would be no NV⁻ centers!), have been fully annealed out during the detonation reaction.

**CONCLUSION**

In this paper, the anomalous formation of the negatively charged NV⁻ centers under electron irradiation in 5 nm sized detonation nanodiamonds (DNDs) was discussed. For the quantification of the NV⁻ centers, the half-field EPR spectroscopy technique[37] was chosen, since optical techniques are limited due to the weak photostability (blinking and bleaching) of NV⁻ centers in such small ND crystals.[32] A comparison with a series of milled HPHT NDs showed that, after identical electron irradiation treatment, the concentration of NV⁻ centers (in units of atomic ratio) in DNDs is higher than that in HPHT NDs of 10 or 20 nm in size. This is remarkable, since the concentration of NV⁻ centers in HPHT NDs increases steadily from 10-100 nm, as predicted earlier by Ref. 47. By modeling the annealing process, simulating the migration of a vacancy by a random walk (Monte Carlo simulation), it was shown that the observed effect maybe partly caused by the higher concentration of substitutional nitrogen atoms [$N_S$] in DNDs (ca. 850-1100 ppm) compared to HPHT NDs (ca. 70-100 ppm). While 10 nm NDs with an equal substitutional nitrogen concentration ([$N_S$] = 100 ppm) show four times more NV⁻ centers than 5 nm particles (in agreement with the $r^2$ behavior)[47], the approximately ten times higher substitutional nitrogen concentration [$N_S$] in DNDs turn this to a three times higher concentration of NV⁻ centers with respect to 10 nm NDs. Our experimental result show a five to six times higher NV⁻ center concentration in DNDs compared to 10 nm HPHT NDs. The remaining difference between the simulated and the experimental value is assigned to the flake-like structure of milled HPHT nanocrystals, effectively reducing the average distance of a vacancy to the surface, and to the omission of charge effects, other defect populations in the diamond lattice and thermodynamic properties in the simulation.

A second anomalous effect in 5 nm DNDs is that the creation of NV$^-$ centers fully takes place during room-temperature irradiation (maximum temperature 100 °C) while the subsequent high-temperature annealing under vacuum (800 °C for 2 h) is not effective.[25] We suggest that a "self-annealing" effect is taking place for DNDs at room temperature (or maximum 100 °C) during annealing. The self-annealing mechanism originates from a combination of "size effects" and "type effects". The small size leads to a low volumetric heat capacity and low heat conductivity, while the high concentrations of substitutional nitrogen atoms [$N_S$] will reduce the necessary diffusion length. If synthetically accessible, 5 nm DNDs with a different concentration in substitutional nitrogen atoms [$N_S$] could experimentally verify this hypothesis. An other experimental verification of our hypothesis, could be in situ local nanodiamond temperature measurements using NV$^-$ thermometry via optically-detected magnetic resonance (ODMR)[57], or using an all-optical approach[58] while irradiating electrons in an electron microscope.

We showed that DNDs can be enriched by NV$^-$ centers from ca. 1 NV$^-$ center in 1000 DND particles (pristine DNDs) to ca. 1 NV$^-$ center in 80 DND particles (at a fluence of $1.5 \times 10^{19}$ e$^-$/cm$^2$, before annealing) and that there is no saturation in sight (see Fig. 2a). Importantly, at the current irradiation conditions, no radiation damage of the DND crystals was observed and the π* transition for sp$^2$ carbon at 285 eV in EELS[59] (e.g., caused by graphitization to carbon onions) did not increase after electron irradiation (TEM and EELS in Fig. S7). Also, it was experimentally shown how the size of nanodiamonds (10 - 100 nm) influences their NV$^-$ concentration, when they are irradiated after milling ("bottom-up approach"). The size dependence, predicted by simulation in Ref. 47, was experimentally confirmed for the first time.

Besides for the nitrogen-vacancy and diamond community, our findings could be of interest for the creation of point defects in very small crystals of other semiconductor materials.

## METHODS

### *Nanodiamonds*

5-nm DNDs as colloidal dispersion were generously donated by Professor Osawa, NanoCarbon Research Institute, Ueda (Japan). Water-dispersed 10 nm high-pressure-high-temperature (HPHT) NDs were purchased from Adamas Nanotechnology (USA); 20 nm ("MSY 0-0.03 micron") and 30 nm ("MSY 0-0.05 micron") HPHT ND powder was purchased from Microdiamant AG (Switzerland). 50 nm ND powder was purchased from Tomei Diamond (Japan); 100 nm NDs, ("MICRON MDA+ M0.10)" powder was purchased from Element Six (UK). In prior to any treatment, water-dispersed DNDs and 10 nm HPHT NDs were freeze-dried with freeze dryer, model FDU-1200, manufactured by EYELA  (Japan), using liquid nitrogen.

### *Electron Irradiation*

Dry ND powders were wrapped in an aluminum foil (sample thickness less than 1 cm) and electron irradiation were performed on a water flow cooled copper plate. The temperature during electron irradiation was measured between the aluminum foil and copper plate, which was around 80 $^{\circ}$C. To investigate the annealing effect caused by the electron irradiation on different sizes of NDs, a series of different sizes nanodiamond (5 nm, 10 nm, 20 nm, 30 nm, 50 nm, 100nm) with $5 \times 10^{18}$ e$^{-}$/cm$^2$ fluence of 2 MeV energy were prepared. In addition to the size series, DNDs and 100 nm NDs irradiated at 2 MeV energy with a fluence of $1 \times 10^{18}$ e$^{-}$/cm$^2$, $3 \times 10^{18}$ e$^{-}$/cm$^2$, $1 \times 10^{19}$ e$^{-}$/cm$^2$, $1.5 \times 10^{19}$ e$^{-}$/cm$^2$ as well as DNDs and 100 nm NDs irradiated at 1 MeV energy with a fluence $5 \times 10^{18}$ e$^{-}$/cm$^2$ fluence were also prepared.

### *Boiling Acid Treatment*

25 mg of ND powder was first transferred into 16 mL of excess nitric acid and sulfuric acid mixture of 1 to 3 volume ratio, and ice-bath sonicated (Bioruptor UCS-200TM, Cosmo Bio, Japan) for 10 minutes to achieve a dispersed state (on/off = 30s/30s). The mixture was stirred and heated at 130 °C for 3 days under reflux inside a fume-hood. NDs were then diluted and cleaned with Milli-Q water for three times at 150,000 RCF to remove metal impurities dissolved in the acid with the aid of Beckman Optima Ultra-centrifugation with aTLA-110 rotor (Beckman Coulter, USA), and a TOMY Digital Biology UR-21P handy sonicator (TOMY, Japan). Finally, the NDs were freeze-dried overnight into powder for EPR measurements using aforementioned equipment.

### *Annealing*

To create NV$^{-}$ centers in the ND sample, ND powders were annealed at 800 $^{\circ}$C for 2 h under vacuum ($p < 10^{-6}$ mbar).[27,60] The temperature was increased to 400 $^{\circ}$C for 1 hour and kept for 4 h ("baking"). In the case of annealing at 400 °C, this "baking" step was skipped. Subsequently, the temperature was increased to 800 °C over 11 h, where the sample was annealed for 2 h. The temperature was decreased to 350 °C for 1 hour and then down to room temperature, using a tailor-made annealing oven KHT-30, Kenix (Japan), composed by a turbomolecular drag pump TSU-261, Pfeiffer Vacuum (Germany), which is connected to an electrical oven NHK-170AF, from Nitto Kagaku Co., Ltd. (Japan).

### *EPR Measurement*

All Electron Paramagnetic Resonance (EPR) spectra were measured using X-band ($\nu = 9 - 10$ GHz) ELEXSYS E 500 with an ELE manufactured by Bruker Corp. (USA). EPR data in Fig. 6 were with a ELEXSYS Super High Sensitivity Probehead (ER 4122SHQE), while the other spectra were measured with a standard EPR cavity. Roughly 20 mg of the powder sample was carefully filled into a 4 mm Thin Wall Quartz EPR sample tube from Wilmad-LabGlass SP Scienceware (USA). ND samples were cleaned with boiling acid treatment (see above) to remove iron impurities to reduce the broad EPR signal around $g_{eff} \approx 4.3$[38] overlapping with the characteristic NV$^-$ HF transition at $g = 4.23$.

### *Mathematical Fitting for Background Elimination*

The remaining background of the HF EPR spectrum was eliminated by mathematical means using background correction and estimation in WinEPR software (Bruker, Inc.).[25,61] The raw first derivative CW EPR spectrum was first integrated once using the WinEPR software to obtain the absorption spectrum. The huge differences in the line shape and line width between the remaining Fe$^{3+}$ signal ($g_{eff}$-value from 4.28 to 4.30) and the narrow NV$^-$ ($g$-value of 4.23 ± 0.01, frequency-dependent) and multi-vacancies ($g$-value of 4.00) signal was used to distinguish the signal of interest from the background signal. Subsequently, points were selected manually outside the region of signals of interest which use to define the undesirable background signals. On spectra with good signal-to-noise ratio, the Functions/Extrapolate/Cubic Spline command was then used to extrapolate the hypothetical baseline. Otherwise, the Functions/Extrapolate/Point to Point command was used to manually extrapolate a hypothetical baseline. After defining the baseline, the subtraction command was applied to obtain the baseline-corrected absorption spectrum. It is worth noting that for the broad $NV^-$ line, such as in the case of DNDs, the aforementioned procedures will distort the line shapes, but generally speaking, the line widths, positions, peak to peak intensities and even the double-integrals were in an acceptable range.

### *Transmission Electron Microscopy (TEM)*

A trace amount of sample was added to Milli-Q water and sonicated for 5 minutes to disperse the powder. Next, few drops of supernatant were transferred onto a hydrophobic Gallium (Ga) thin film (thickness of 10 nm) and dried under a heating lamp for 1 hour. The film was then transferred into JIC-410 ion cleaner for weak plasma cleaning to remove any contamination from the air. Finally, the sample on Ga thin film was placed into the JEOL JEM-2000FC Transmission Electron Microscopy of a 0.19 nm point resolution, 200 kV accelerated, ZrO/W (100) FEG (Schottky) electron gun, in-column omega energy spectrometer of 0.8 eV energy resolution. All electron energy loss spectroscopy (EELS) spectra were recorded with the same machine at a circle diameter of 235 nm and an energy resolution of 1.2 eV.

### *Monte Carlo Simulation*

The NV formation probability was calculated in a computational method using Wolfram Mathematica version 11.3.0 with a vacancy moving in a generated diamond crystal structure. The preparation of the method consists of the following two steps: the crystal structure generation and replacement of carbon with nitrogen atoms. In the first step, since the coordinate of carbon atoms in the unit cell of a diamond are $(0, 0, 0)$, $(a/2, a/2, 0)$, $(a/2, 0, a/2)$, $(0, a/2, a/2)$, $(a/4, a/4, a/4)$, $(3a/4, 3a/4, a/4)$, $(3a/4, a/4, 3a/4)$ and $(a/4, 3a/4, 3a/4)$, where a is the lattice constant, a crystal structure was generated by repeating the unit cell. The generated cubic crystal structure was reshaped into a spherical structure with the radius of interest. Each atom was associated with the component information on a unique number, the unique numbers of adjacent atoms, the coordinate, and a property including carbon, nitrogen, and edge. The edge property identifies the atoms on the surface of the spherical crystal structure, corresponding to the atoms whose number of the unique numbers of adjacent atoms is not four. In the second step, the carbon atoms were replaced with nitrogen atoms by replacing the property in the component information.

The NV$^-$ formation probability calculation consists of the unit process with replacement with a vacancy, vacancy movement, and termination. The unit process was repeated $10^4$ times to calculate the ratio of the creation of a nitrogen vacancy center, reaching a conversion. First of all, a carbon atom in the generated spherical crystal structure was randomly selected as a vacancy. The vacancy was moved to an adjacent atom randomly selected from one of the numbers of adjacent atoms in the component information. This is repeated until the vacancy reaches the edge of the crystal ("failure") or an adjacent lattice site to a nitrogen atom ("success"). The NV formation probability was calculated as the ratio of the "success" cases divided by the total number of runs.

**Supporting Information Available**: All EPR spectra (raw data), further Monte-Carlo simulation data, TEM images with EELS spectra, DLS size distribution and protocols for estimation of P1 content in DNDs and small NDs as well as the content of isolated P1 centers in larger diamond particles. The Mathematica script for the Monte Carlo simulation will be shared after contacting the corresponding authors.

This material is available free of charge via the Internet at [http://pubs.acs.org](http://pubs.acs.org). Supporting Information (PDF)

## Author Contributions

The manuscript was written through contributions of all authors. All authors have given approval to the final version of the manuscript.

## Funding Sources

- The Branco Weiss Fellowship - Society in Science

- MEXT Quantum Leap Flagship Program (MEXT Q-LEAP) (JPMXS0120330644 and JPMXS0118067395)

- Japan Society for the Promotion of Science, KAKENHI (Grants 20H00453, 18K19297, 21H04444)

## ACKNOWLEDGEMENT

T.F.S. thanks Amar Vutha (University of Toronto, Canada) for fruitful discussions with many helpful insights. The authors thank Taras Plakhotnik (University of Queensland, Australia) for kindly sharing the original simulation data in Fig. 6. T.F.S. acknowledges The Branco Weiss Fellowship – Society in Science, administrated by the ETH Zurich, and Prof. Shirakawa for hosting him as a Guest Research Associate at Kyoto University. We gratefully acknowledge financial support in the MEXT Quantum Leap Flagship Program (MEXT Q-LEAP) (JPMXS0120330644 and JPMXS0118067395) and Japan Society for the Promotion of Science, KAKENHI (Grants 20H00453, 18K19297, 21H04444). The TEM imaging-related work was supported by Kyoto University Nano Technology Hub in "Nanotechnology Platform Project" sponsored by the Ministry of Education, Culture, Sports, Science and Technology (MEXT), Japan. F.T.-K.S. acknowledges the Asian Future Leaders Scholarship Program, administrated by Bai Xian Asia Institute (BXAI) and Kyoto University for support his Master degree during this project.

## ABBREVIATIONS

DNDs, detonation nanodiamonds; EELS, electron energy loss spectroscopy; EPR, electron paramagnetic resonance; HF transition, half-field transition; HPHT, high-pressure-high-temperature; NDs, nanodiamonds; $NV^-$ center, negatively charged nitrogen-vacancy center; PL, photoluminescence; SAED, selected-area electron diffraction; TEM, transmission electron microscopy.

# Supplementary Information

# The Anomalous Formation of Irradiation Induced Nitrogen-Vacancy Centers in 5-Nanometer-Sized Detonation Nanodiamonds

Frederick T.-K. So[1,2,3], Alexander I. Shames[4], Daiki Terada[2,3], Takuya Genjo[2,3], Hiroki Morishita[1], Izuru Ohki[1], Takeshi Ohshima[5], Shinobu Onoda[5], Hideaki Takashima[6], Shigeki Takeuchi[6], Norikazu Mizuochi[1], Ryuji Igarashi[3,7], Masahiro Shirakawa[2,3]*, Takuya F. Segawa[2,8]*

[1]Institute for Chemical Research, Kyoto University, Gokasho, Uji, Kyoto, 611-0011, Japan.

[2]Department of Molecular Engineering, Graduate School of Engineering, Kyoto University, Nishikyo-Ku, Kyoto 615-8510, Japan

[3]Institute for Quantum Life Science, National Institutes for Quantum Science and Technology, Anagawa 4-9-1, Inage-ku, Chiba 263-8555, Japan

[4]Department of Physics, Ben-Gurion University of the Negev, Israel

[5]Takasaki Advanced Radiation Research Institute, National Institutes for Quantum Science and Technology, 1233 Watanuki, Takasaki, Gunma 370-1292, Japan

[6]Department of Electronic Science and Engineering, Kyoto University, Kyotodaigakukatsura, Nishikyo-ku, Kyoto 615-8510, Japan

[7]National Institute for Radiological Sciences, National Institutes for Quantum Science and Technology, Anagawa 4-9-1, Inage-ku, Chiba 263-8555, Japan

[8]Laboratory for Solid State Physics, ETH Zurich, Otto-Stern-Weg 1, 8093 Zürich, Switzerland

* E-mail: shirakawa@moleng.kyoto-u.ac.jp; takuya.segawa@alumni.ethz.ch

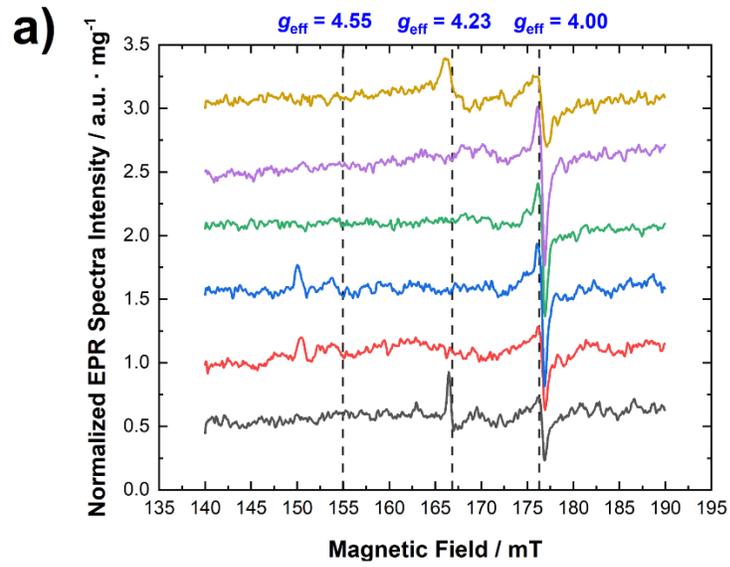

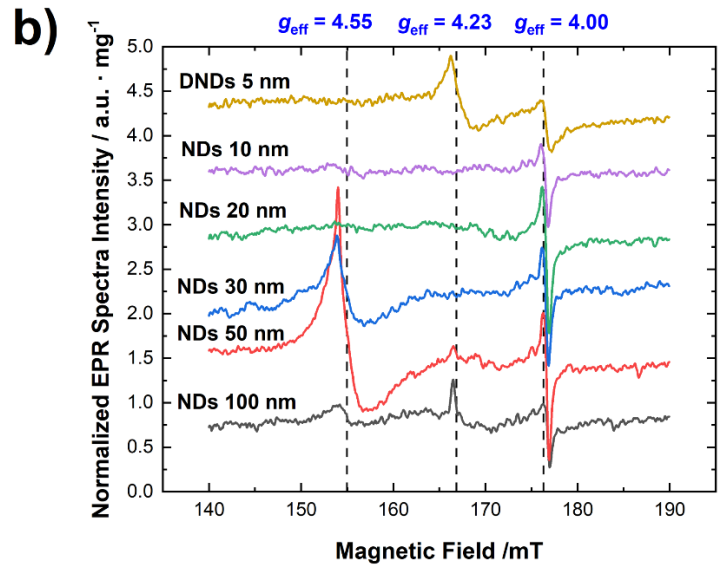

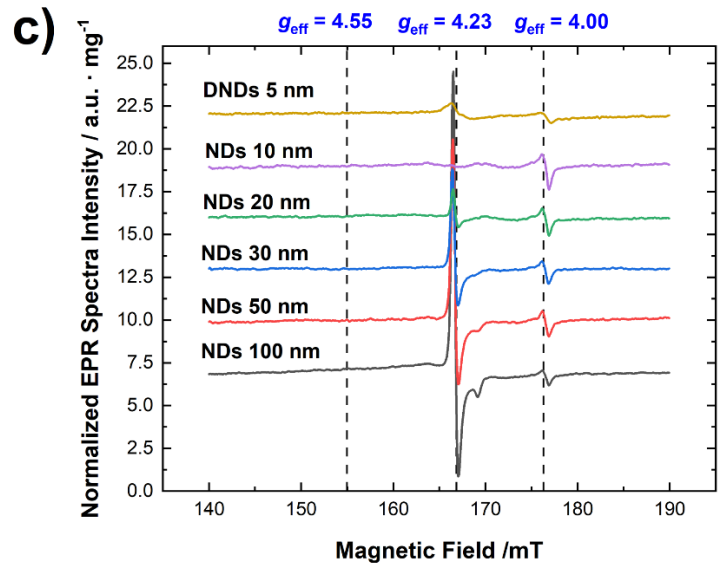

Figure S1. Half-field EPR spectra of electron-irradiated (2 MeV, $5 \times 10^{18}$ e$^-$/cm$^2$) nanodiamonds with different particle size, $\nu = 9.87^*$ GHz. a) pristine; b) electron irradiated; c) electron irradiated with annealing. These are the EPR spectra for Fig. 1 (blue, orange, and yellow bars), where the double integral of the signal at $g_{eff} = 4.23$ leads to the NV$^-$ concentration. The signal at $g_{eff} = 4.55$ are so-called R1/R2 defects[1,2], which disappear after annealing. The signal at $g_{eff} = 4.00$ is attributed to multi-vacancy defects. Under the same electron-irradiation condition (2 MeV, $5 \times 10^{18}$ e$^-$/cm$^2$), there were almost no for R1/R2 defects (two types of self-interstitial defects) for 5, 10, 20 nm NDs, a strong R1/R2 signal for 30 and 50 nm NDs, and again a weak R1/R2 signal for 100 nm NDs.

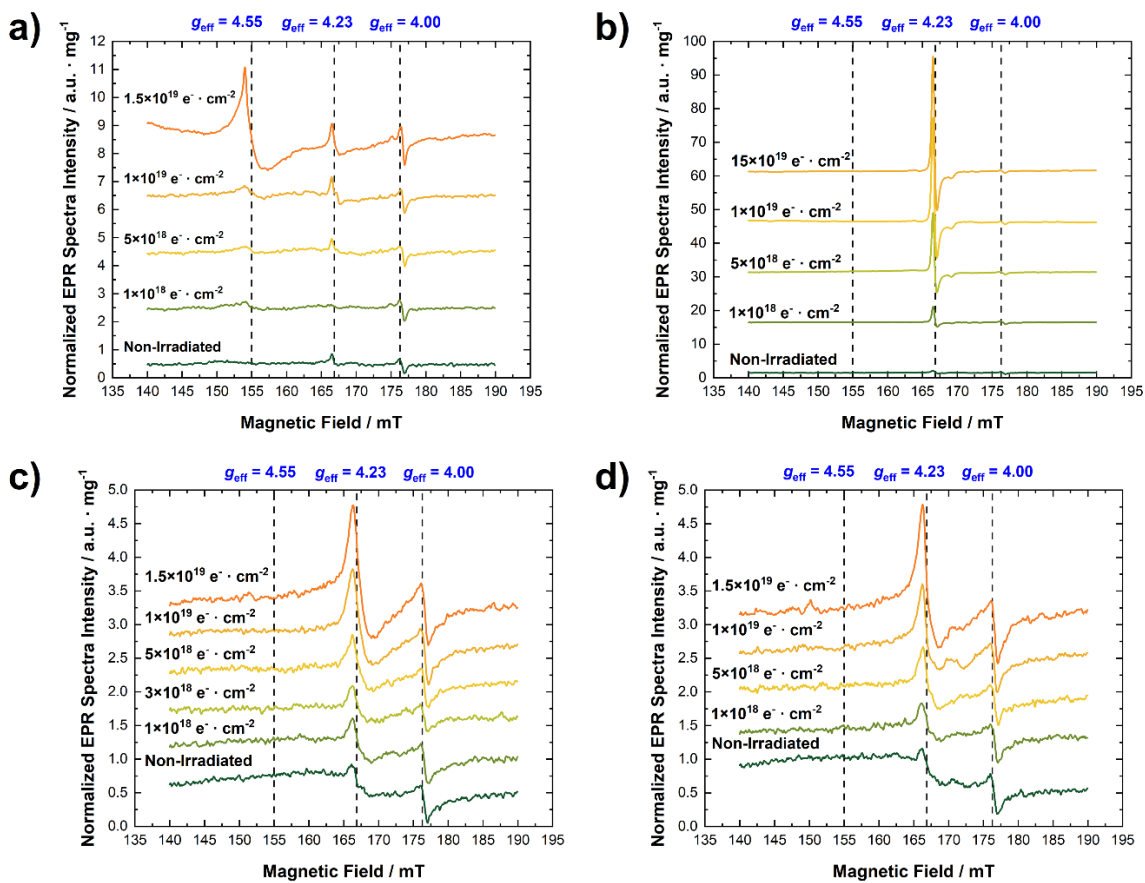

Figure S2. Half-field EPR spectra of electron-irradiated (2 MeV) nanodiamonds with different electron fluence, ν = 9.87* GHz. (a) electron-irradiated 100 nm HPHT NDs; (b) electron-irradiated annealed (800 °C) 100 nm HPHT NDs; c) electron-irradiated 5 nm DNDs; (d) electron-irradiated annealed (800 °C) 5 nm DNDs. These are the EPR spectra for Fig. 2, where the double integral of the signal at $g_{eff}$ = 4.23 leads to the NV⁻ concentration. Upon increasing the electron irradiation fluence, 100 nm HPHT NDs showed an increasing intensity of R1/R2 defects, which completely disappeared after annealing at 800 °C for 2 h. Notably, DNDs showed no observable signal of R1/R2 after electron irradiation.

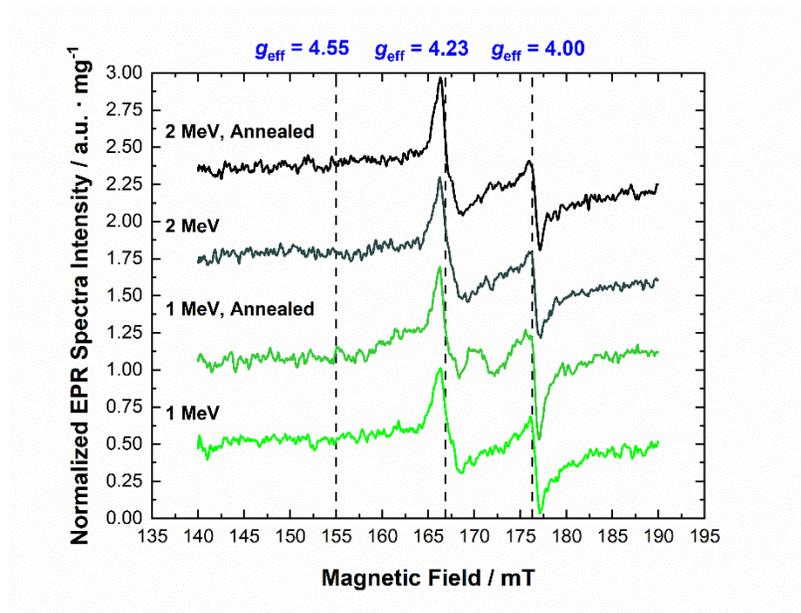

Figure S3. Half-field EPR spectra of 5 nm DNDs with different electron irradiation energy (2 MeV or 1 MeV), with or without annealing, $\nu$ = 9.87* GHz. These data are the EPR spectra (before baseline correction) of Fig. 3 in the main text.

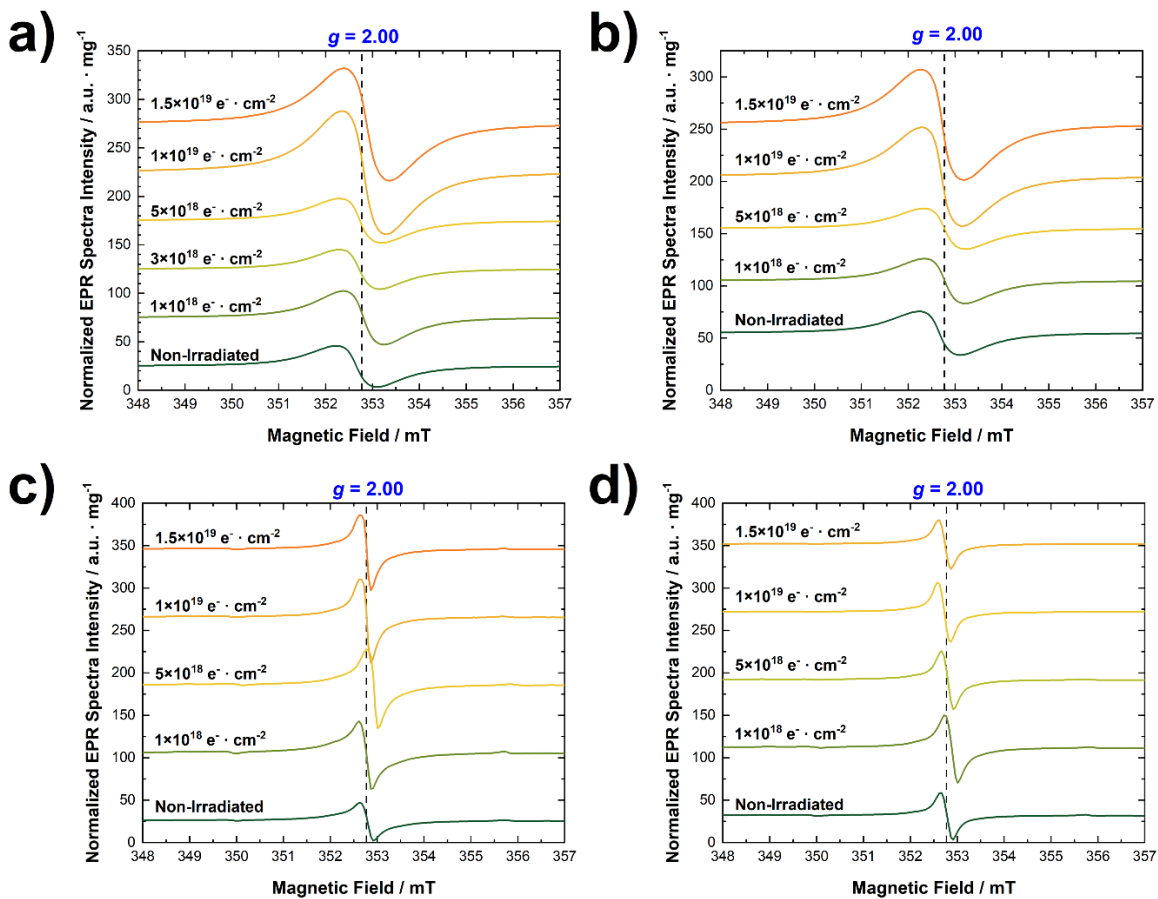

Figure S4. Main-field EPR spectra ($g \approx 2$) of the data in Fig. 4, $\nu = 9.87*$ GHz; (a) electron-irradiated DNDs; (b) electron-irradiated and annealed DNDs; (c) electron-irradiated 100 nm HPHT NDs; (d) electron-irradiated and annealed 100 nm HPHT NDs. Electron irradiation was performed at temperature below 100 ˚C in air while annealing was performed at 800 ˚C in vacuum for 2 h.

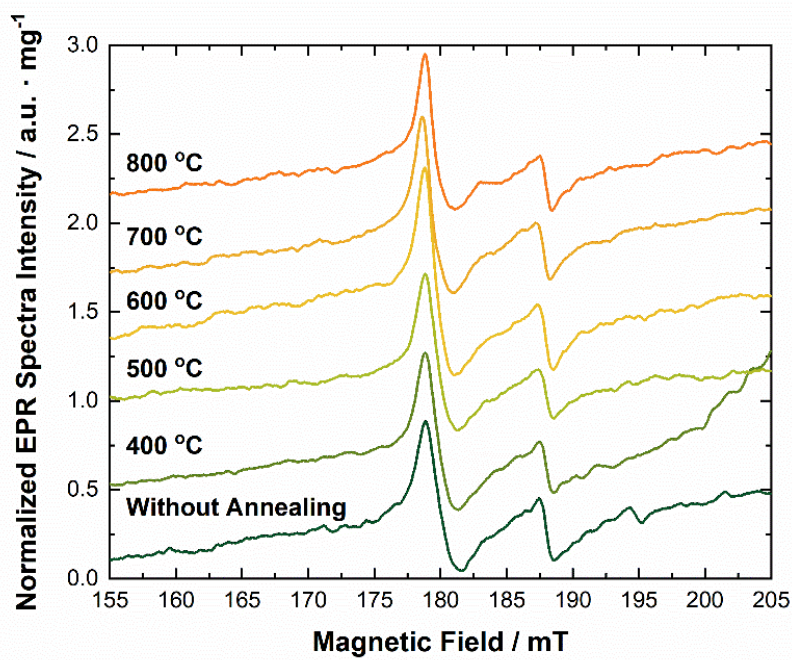

Figure S5. Half-field EPR spectra for the data in Fig. 5, $\nu$ = 9.87* GHz; electron-irradiated ($1.5 \times 10^{19}$ e⁻/cm², 2 MeV) DNDs as a function of different annealing temperatures, ranging from 800 °C down to 400 °C (for 2 h) and without annealing.

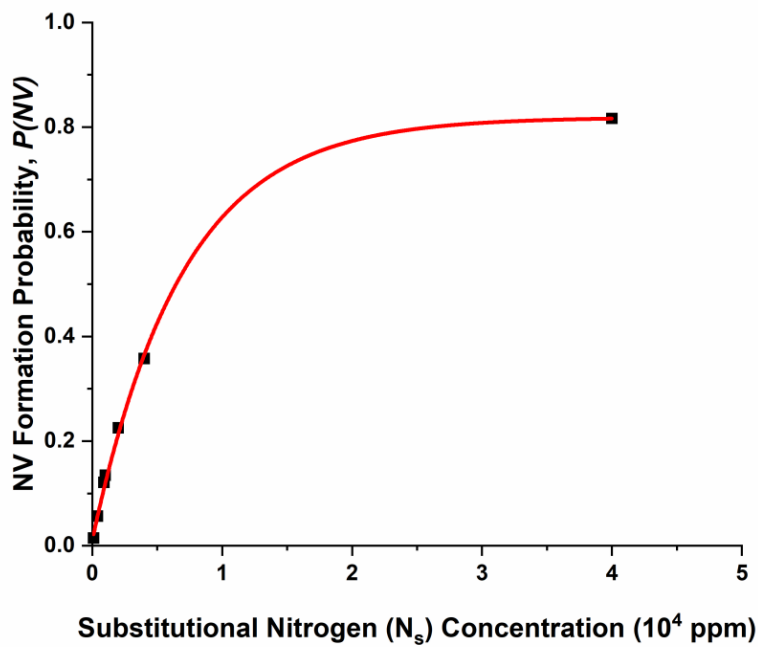

Figure S6. Result of Monte Carlo annealing simulation, plotting the NV formation probability as a function of the substitutional nitrogen concentration [$N_S$] of a 5 nm-sized spherical diamond crystal (black squares). Upon increasing the substitutional nitrogen concentration [$N_S$], the NV formation probability showed a saturating behavior, which was fitted (red line) with an exponential buildup and a constant of 7000 ppm (in units of atomic ratio). The first five simulated values up to [$N_S$] = 1000 ppm correspond to the data points for a particle diameter of 5 nm in Fig. 7.

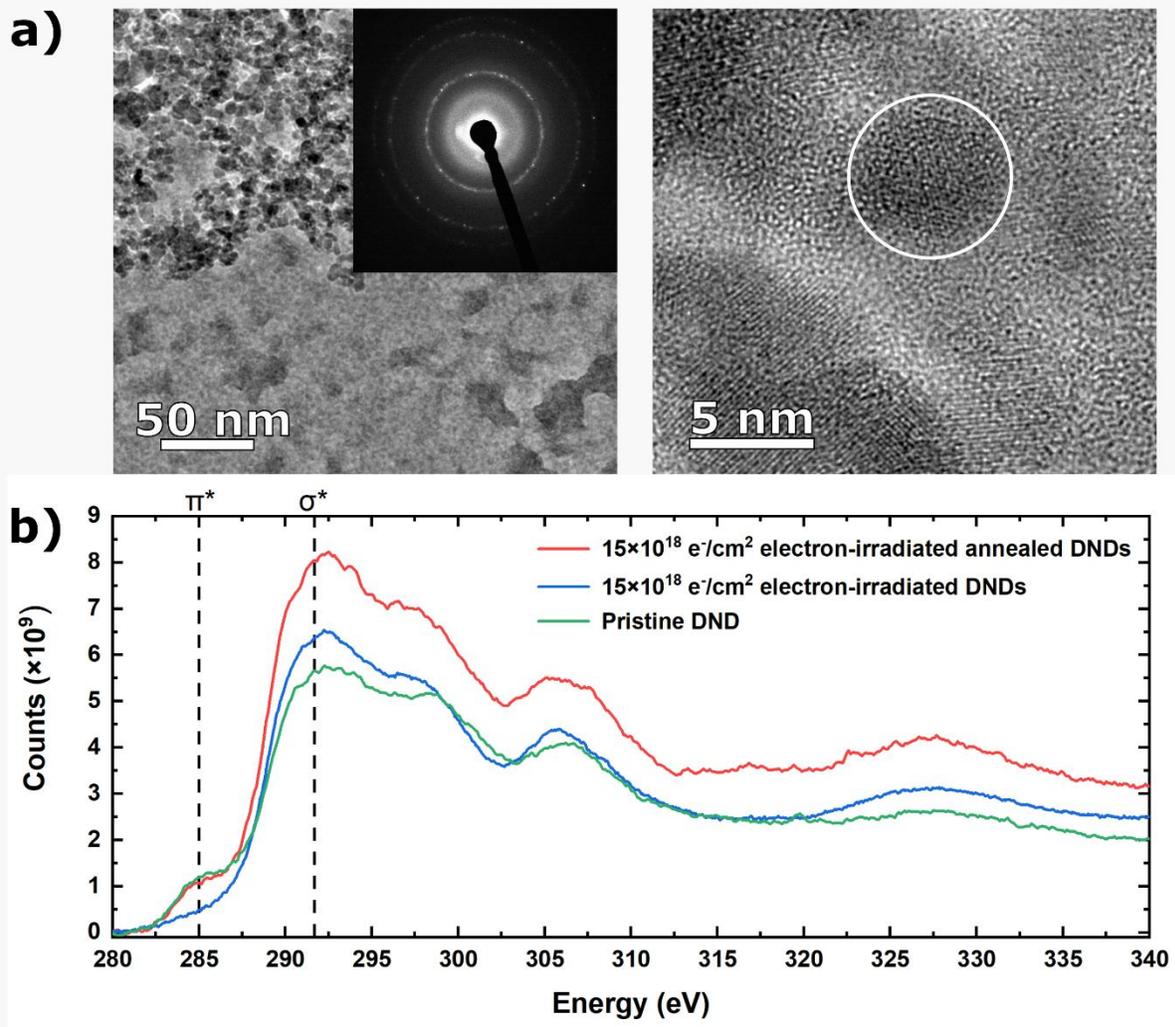

Figure S7. Characterization of $1.5 \times 10^{19}$ e$^-$/cm$^2$ electron-irradiated and annealed DNDs. (a) Representative TEM image of a single 5 nm DND. The inset shows selected-area electron diffraction (SAED). (b) Electron energy loss spectroscopy (EELS) of DNDs: $1.5 \times 10^{19}$ e$^-$/cm$^2$ irradiated and annealed DNDs (red); $1.5 \times 10^{19}$ e$^-$/cm$^2$ irradiated DNDs (blue); pristine DNDs (green). The energy at 285 eV corresponds to the π* transition of sp$^2$ carbon, 291.7 eV to the σ* transition stemming from sp$^3$ carbon in diamond.[3]

To investigate the structural integrity of DNDs after the highest electron irradiation in our work ($1.5 \times 10^{19}$ e$^-$/cm$^2$ with 2 MeV electrons), the samples were imaged under TEM. DND crystallites with sizes of 5 nm (Fig. S6a) and corresponding SAED with the typical parameters from the diamond lattice were recorded with no visible changes to pristine DND samples (see Ref. 4). To investigate on irradiation-induced graphitization, EELS spectra were additionally recorded (Fig. S6b). The π* transition for $sp^2$ carbon at 285 eV[3] did not increase after electron irradiation. Please note that the height of the peaks depends on the number of particles in the area of observation. The slight increase at 285 eV π* transition for the "Pristine DND" (green) and the "Annealed DND" (red) is most likely contamination from the sample holder.

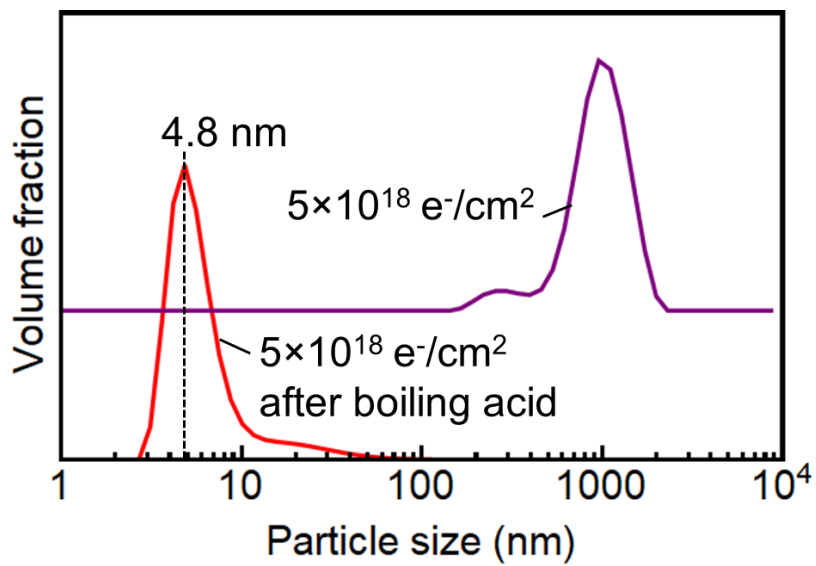

Figure S8. Size distribution of DND particles measured with dynamic light scattering (DLS). $5 \times 10^{18}$ e⁻/cm² electron-irradiated DNDs before (purple) and after the boiling acid treatment (red). This measurements confirms the deaggregated particle size of ca. 5 nm. Replotted from Ref. 4 (Fig. 3(a) therein).

# Estimation of P1 content in DNDs and small NDs as well as the content of isolated P1 centers in larger diamond particles

The DND samples under study were obtained from NanoCarbon Research Institute Ltd. (Nagano, Japan). Two samples were from different batches of freeze-dried aqueous colloidal solutions of 5-nm bead-assisted sonication disintegrated (BASD) DNDs. The third DND sample was freeze-dried boiling acid cleaned sample. The reference purified DND sample with the known density of $S = 1/2$ paramagnetic centers $N_S = 6.3\times10^{19}$ spin/g and $g = 2.0028 \pm 0.0002$[5] was obtained from Ioffe Institute (St. Petersburg, Russia)[5].

Continuous wave X-band (9.4 GHz) EPR measurements were carried out using a Bruker EMX - 220 spectrometer (Bruker Scientific Israel Ltd., Rehovot, Israel) installed at the Ben-Gurion University of the Negev (Israel) and equipped with a 53150A frequency counter (Agilent Technologies Inc., Santa Clara, CA) and an Oxford Instruments ESR900 variable temperature accessory (Oxford Instruments plc, Tubney Woods, UK). The spectra were acquired at room temperature (RT, $T \sim 295$ K). Accurate determination of electronic $g$-factors and densities of paramagnetic $S = 1/2$ species were done against the aforementioned reference DND sample. EPR data processing and simulation were carried out using WIN-EPR (Bruker) and OriginLab (OriginLab Corporation, Northampton, MA) software packages.

15 – 30 mg of each DND powder (having different bulk densities) were placed into EPR silent 4 mm o. d. Wilmad quartz tube in attempt to keep the same filling factor ($\sim 15$ mm height) and centered into the Bruker ER 4102ST standard rectangular cavity. The Q-factor of loaded cavity varied within the range of 3700 – 4300. General view (scan width SW = 0.66 T) EPR spectra were recorded at microwave power $P_{MW} = 20$ mW, 100 kHz magnetic field modulation amplitude $A_m = 1$ mT, receiver gain RG = $2\times10^5$, number of coherent acquisition $n_{aq} = 1$, digital resolution N =

2048 points and microwave frequency ν = 9.466 GHz. The general view spectra (not show) revealed intense singlet signal with $g \sim 2.00$, low intensity broad signals with $g_{eff} \sim 4.3$ as well as weak doublet of relatively narrow signals with $g_{eff} = 4.26$ and $4.00$. Boiling acid cleaned BASD DND sample as well as the reference purified DND sample showed significantly reduced $g_{eff} \sim 4.3$ signals. High resolution EPR spectra of the intense $g \sim 2.00$ signal were recorded at SW = 0.04 T, non-saturating $P_{MW} = 0.2$ mW, $A_m = 0.02$ mT, RG = $2 \times 10^4$ and $1 \times 10^5$, $n_{aq} = 1$ and $n_{aq} = 100$, N = 2048 points and microwave frequency ν = 9.466 GHz. Black open circles in Fig. S8 show the experimental EPR spectrum of the boiling acid treated BASD DND sample. The signals observed in the DND samples under study are typical for DNDs (see, for instance Ref. 5) and may be described as a singlet line having quasi-Lorentzian line shape, $g_{iso} = 2.0028 \pm 0.0002$ and line width $\Delta H_{pp}$ varying within the range of $0.85 - 0.90$ mT (depending on sample's purity). The average (over three samples and two independent measurements) density of all $S = 1/2$ spins was found $N_S = (6.4 \pm 0.9) \times 10^{19}$ spin/g, or $1276 \pm 220$ ppm.

It was recently found that the singlet quasi-Lorentzian-like line, observed in DNDs may be successfully decomposed into two Lorentzian lines (narrow and broad) having very close $g$-factor values.[6] The applicability of the Lorentzian line shape model for the description of individual line shapes of both surface (2D) and bulk (3D) dipole-dipole and exchange coupled paramagnetic centers in carbon containing materials has been carefully studied and convincingly confirmed by Atsarkin et al.[7] We applied the aforementioned two-component model to simulating the singlet lines in the DND samples under study. The experimental EPR signals (first derivative of the EPR absorption signals) were fitted using the following equation

$$Y'(H) \propto \sum_{i=\text{ narrow, broad}} C_i \left\{ \frac{\left(\dfrac{H - H_{\tau i}}{\Delta H_{\text{pp}i}}\right)}{\left[3 + 4\left(\dfrac{H - H_{\tau i}}{\Delta H_{\text{pp}i}}\right)^2\right]^2} + \frac{\left(\dfrac{H + H_{\tau i}}{\Delta H_{\text{pp}i}}\right)}{\left[3 + 4\left(\dfrac{H + H_{\tau i}}{\Delta H_{\text{pp}i}}\right)^2\right]^2} \right\}, \qquad (1)$$

where $H_{\tau i}$ is the resonance field of the corresponding (broad or narrow) signal and $\Delta H_{\text{pp}i}$ is its line width. The use of the two terms in Eq. (1) accounting for the clockwise as well as the anticlockwise circularly polarized component of the microwave radiation is necessary for the precise simulation of the broad Lorentzian lines.[8] Fig. S8 shows the results of fitting the experimental spectrum (black open circles in Fig. S8) by the sum of two Lorentzian lines (red solid line in Fig. S8) with the same (within the experimental error) resonance fields ($g$-factors) – narrow (blue solid line) and broad (green solid line) ones. It is clearly seen that this model describes the experimental line shape quite satisfactorily ($R^2 = 0.998$).

Using transition metal ions ($Cu^{2+}$, $Gd^{3+}$) grafted to DND as paramagnetic probes allowed to ascertain that broad and narrow Lorentzian components belong to different types of $S = 1/2$ defects distinguishing by their location inside the DND particle.[6,9,10] Thus, it was established that defects responsible for the broad Lorentzian component are located at distances of ~0.8 nm from the DND surface whereas defects providing the narrow Lorentzian component are located deeper – at distances of ~1.5 nm from the surface.[10] Further step towards the correct attribution of these two groups of defects was done in the detailed study of the evolution of EPR spectra of HPHT diamonds on diminishing the average particles' size from hundreds micrometers to tens nanometers.[11] EPR spectra showed consecutive transformation of the well resolved hyperfine pattern due to paramagnetic substitutional nitrogen centers (P1 centers with $S = 1/2$ coupled to $^{14}N$ nuclear spins with $I = 1$), observed in spectra of large micron sized diamonds, to the singlet quasi-Lorentzian signal in the spectra of small nanodiamonds. It was supposed that the broad Lorentzian component originates from dangling bonds, located within the surface/interface layers of the nanodiamond particle, whereas the narrow component originates from the initial P1 defects which content remains practically unchanged on micronization and further fractionation, but their hyperfine spectrum turned to a Lorentzian line due to strong exchange interaction via surface defects.[11,12] Close similarity of the EPR spectra of DNDs and the same spectra of nitrogen containing HPHT diamonds micronized to small nanoparticles allows attributing the aforementioned model of singlet double component quasi-Lorentzian signals to DNDs. Thus, one may suppose that the broader component belongs to dangling bonds and the narrow component belongs to P1 centers. This model allows reliable estimation of the content of P1 centers in DND. Indeed, decomposition into two Lorentzians provides data on relative contribution of each group of $S = 1/2$ defects into the known total content of defects, responsible for the intense EPR signal in DND. It was found that for each DND samples under study the narrow signal contributes $(76 \pm 3)\%$ (averaged over three samples) of the total spins, which corresponds to the P1 content of $964 \pm 110$ ppm.

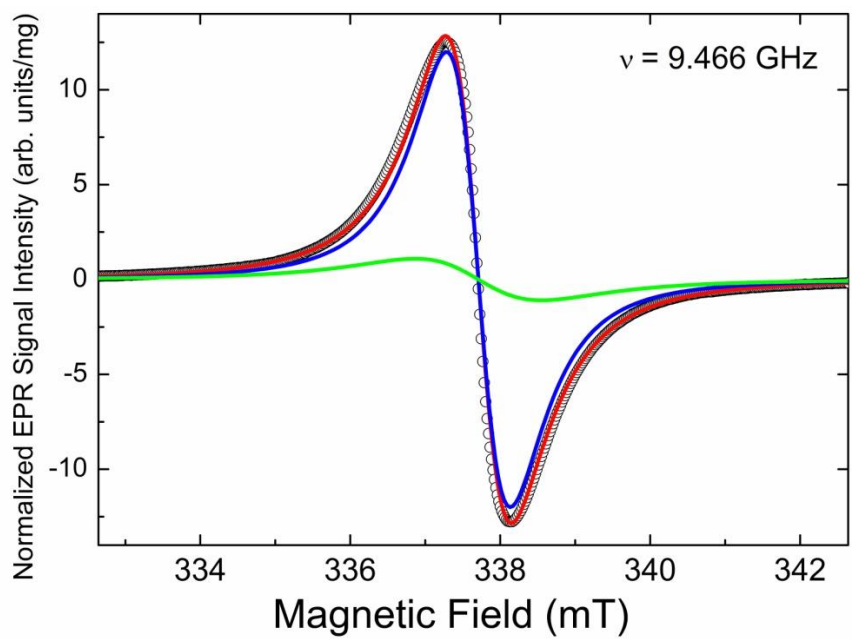

Figure S9. Room temperature EPR spectra of the freeze-dried boiling acid cleaned BASD DND sample recorded at SW = 0.04 T, non-saturating $P_{MW}$ = 0.2 mW, $A_m$ = 0.02 mT, RG = 1×10$^5$, $n_{aq}$ = 100, N = 2048 points and ν = 9.466 GHz: black open circles – experimental spectrum, red solid line – best least square fit by two Lorentzian components, blue solid line – narrow Lorentzian component, green solid line – broad Lorentzian component.

In contrast to the aforementioned procedure, the estimation of the content of isolated (weakly interacting) P1 defects in larger NDs both initial and undergone irradiation/annealing treatment is more obvious procedure. Thus, in initial (not irradiated) large micron sized HPHT diamond particles isolated P1 centers are primary paramagnetic centers. Their RT EPR spectrum within the $g$ = 2.00 region contains only weak contribution of other paramagnetic centers (exchange coupled P1, dangling bonds, P2 centers).[1] In these cases the simulated spectrum of P1 centers is in a good agreement with the experimental spectrum. Therefore direct double integration of the entire experimental spectrum against the same done for the reference sample provides reliable P1 content. However, both reduction of particle size and irradiation/annealing create noticeable amounts of other paramagnetic defects, which makes the EPR spectrum quite different from the "pure" isolated P1 spectrum. Fig. S9(a) represents experimental RT EPR spectrum of e-beam irradiated (fluence 3×10$^{18}$ e$^-$/cm$^2$) and annealed type Ib HPHT diamond sample crushed to particles of 100 nm average size. Simulating typical P1 pattern (in absorption mode) reveals that isolated P1 particles just partially contribute into the experimental spectrum – compare red solid and black dashed lines in Fig. S9(b).

---

[1] Here is worth mentioning that the EPR signal due to substitutional nickel defects (Ni$_s^-$) which line has $g$ = 2.032 (i.e. in the low field wing of the polycrystalline P1 pattern) is observable only at temperatures below 150 K.

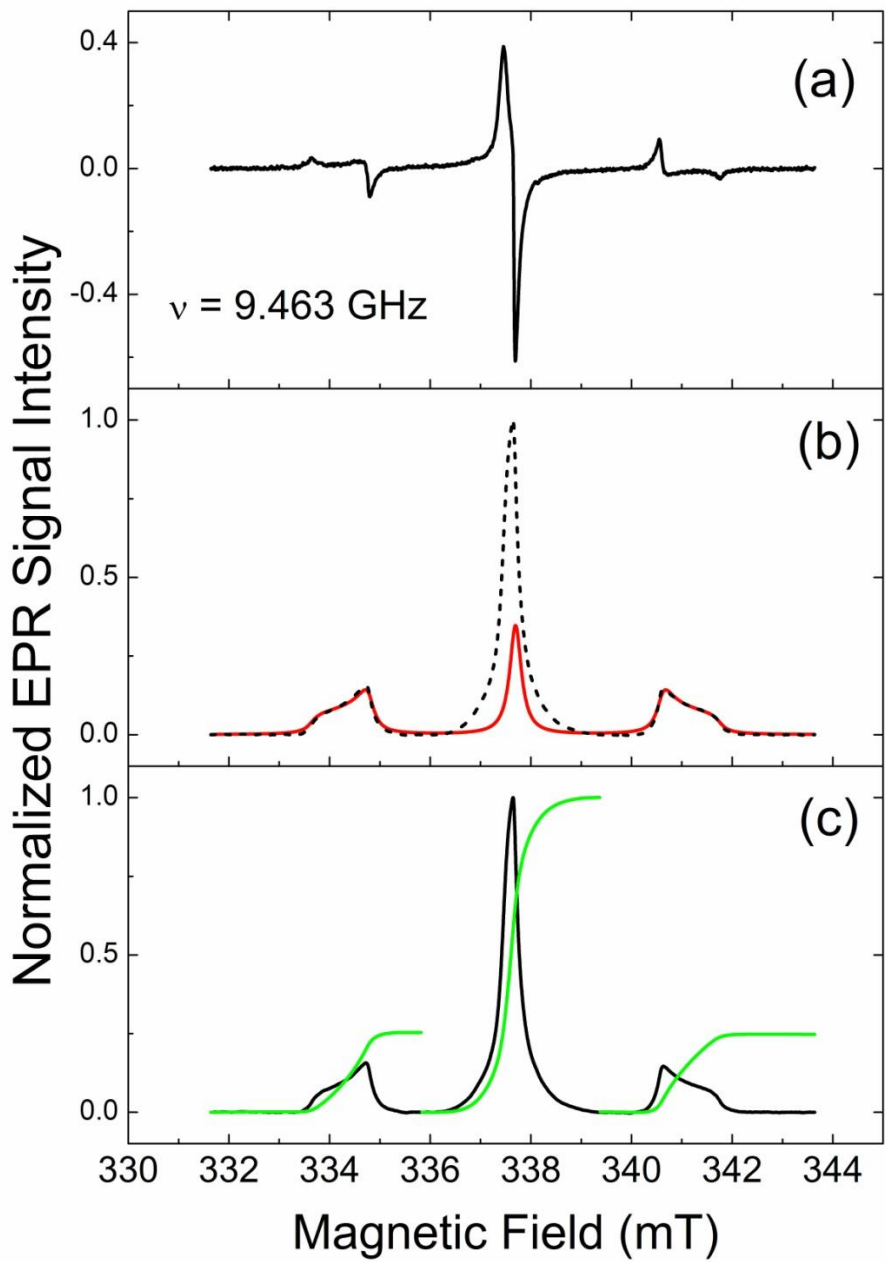

Figure S10. Room temperature EPR spectra of  100 nm average particle size $3\times10^{18}$ e⁻/cm² irradiated/annealed HPHT diamond sample recorded at SW = 0.012 T, non-saturating $P_{MW}$ = 0.002 mW, $A_m$ = 0.01 mT, RG = $2\times10^5$, $n_{aq}$ = 100, N = 2048 points and ν = 9.463 GHz: (a) experimental

spectrum; (b) black dashed line - experimental spectrum, absorption signal, red solid line – simulated absorption signal for P1 centers, simulation parameters are $A_{xx} = A_{yy} = 2.9$ mT, $A_{zz} = 4.058$ mT, $g_{iso} = 2.0024$, $\Delta H_{pp}(L) = 0.2$ mT; (c) black solid line - experimental spectrum, absorption signal, green solid lines – integrated components of the triplet hyperfine absorption pattern.

The main difference is due to the central component which, in addition to $m_I = 0$ P1 hyperfine line, involves the presence of EPR signatures of other paramagnetic defects with $g$-factors within the region $2.0032 – 2.0023$. The difference between the integrated values of the entire central component and predicted integrated value of the $m_I = 0$ P1 hyperfine line (determined as the average values measured for well resolved low- and high-field hyperfine components) occurs just due to other paramagnetic centers. Fig. S9(c) clearly demonstrates this observation. It allows to estimate the content of isolated P1 centers by calculating the contribution of triplicated value of a single hyperfine component into the total spin content. Thus, for the aforementioned irradiated/annealed 100 nm sample having total content of $S = 1/2$ defects of 106 ppm, the isolated P1 centers contribute 50%, which results in 53 ppm.